\documentclass[journal=jpcbfk,manuscript=article]{achemso}

\usepackage[version=3]{mhchem} 
\usepackage[T1]{fontenc}       
\usepackage{amsmath}

\usepackage[usenames, dvipsnames]{color}
\SectionNumbersOn
\usepackage[section]{placeins}
\usepackage{tabularx}
\usepackage{graphicx}
\usepackage{lscape}
\usepackage{subfigure}
\usepackage{epstopdf}
\usepackage{bm}
\usepackage{float}
\usepackage{dblfloatfix}
\setkeys{acs}{usetitle = true}
\usepackage{gensymb}
\usepackage[version=3]{mhchem}
\usepackage{fixltx2e}
\usepackage{tikz}
\usepackage{pdfpages}

\newcommand{\cforty}{C$_{45}$}

\newcommand{\pvnt}{P_v(\tilde{N})}

\newcommand{\gvnt}{G_v(\tilde{N})}

\newcommand{\Uphi}{U_\phi}
\newcommand{\Nt}{\tilde{N}}
\newcommand{\Ntv}{\tilde{N}_v}

\newcommand{\avgNt}{\langle \tilde{N_v} \rangle}
\newcommand{\avgNtphi}{\langle \tilde{N_v} \rangle_\phi}
\newcommand{\varNtphi}{\langle \delta \tilde N_v^2 \rangle_\phi}
\newcommand{\Rg}{R_{\rm g}}

\newcommand{\kbt}{k_{\rm{B}} T}

\newcommand{\Nte}{\Nt_{\rm E}}
\newcommand{\Ntc}{\Nt_{\rm C}}

\newcommand{\one}{ {\large \raisebox{0.5pt}{\textcircled{\raisebox{-.9pt} {\small I}}}} }
\newcommand{\two}{ {\large \raisebox{0.5pt}{\textcircled{\raisebox{-.9pt} {\small II}}}} }
\newcommand{\three}{ {\large \raisebox{0.5pt}{\textcircled{\raisebox{-.9pt} {\small III}}}} }

\usetikzlibrary{calc}

\let\oldmaketitle\maketitle
\let\maketitle\relax
\usepackage{placeins}

\let\Oldsection\section
\renewcommand{\section}{\FloatBarrier\Oldsection}

\let\Oldsubsection\subsection
\renewcommand{\subsection}{\FloatBarrier\Oldsubsection}
\author{Debdas Dhabal}
\affiliation{Department of Chemical and Biomolecular Engineering, University of Pennsylvania, Philadelphia, Pennsylvania 19104, United States}
\author{Zhitong Jiang}
\affiliation{Department of Chemical and Biomolecular Engineering, University of Pennsylvania, Philadelphia, Pennsylvania 19104, United States}
\author{Amish J. Patel}
\email{amish.patel@seas.upenn.edu}
\affiliation{Department of Chemical and Biomolecular Engineering, University of Pennsylvania, Philadelphia, Pennsylvania 19104, United States}

\title[An \textsf{achemso} demo]
  {Characterizing the Interplay between Polymer Solvation and Conformation}


\begin{document}
\setlength{\fboxrule}{0 pt}
\begin{tocentry}
\begin{center}
\includegraphics[clip=true,width=12.0cm]{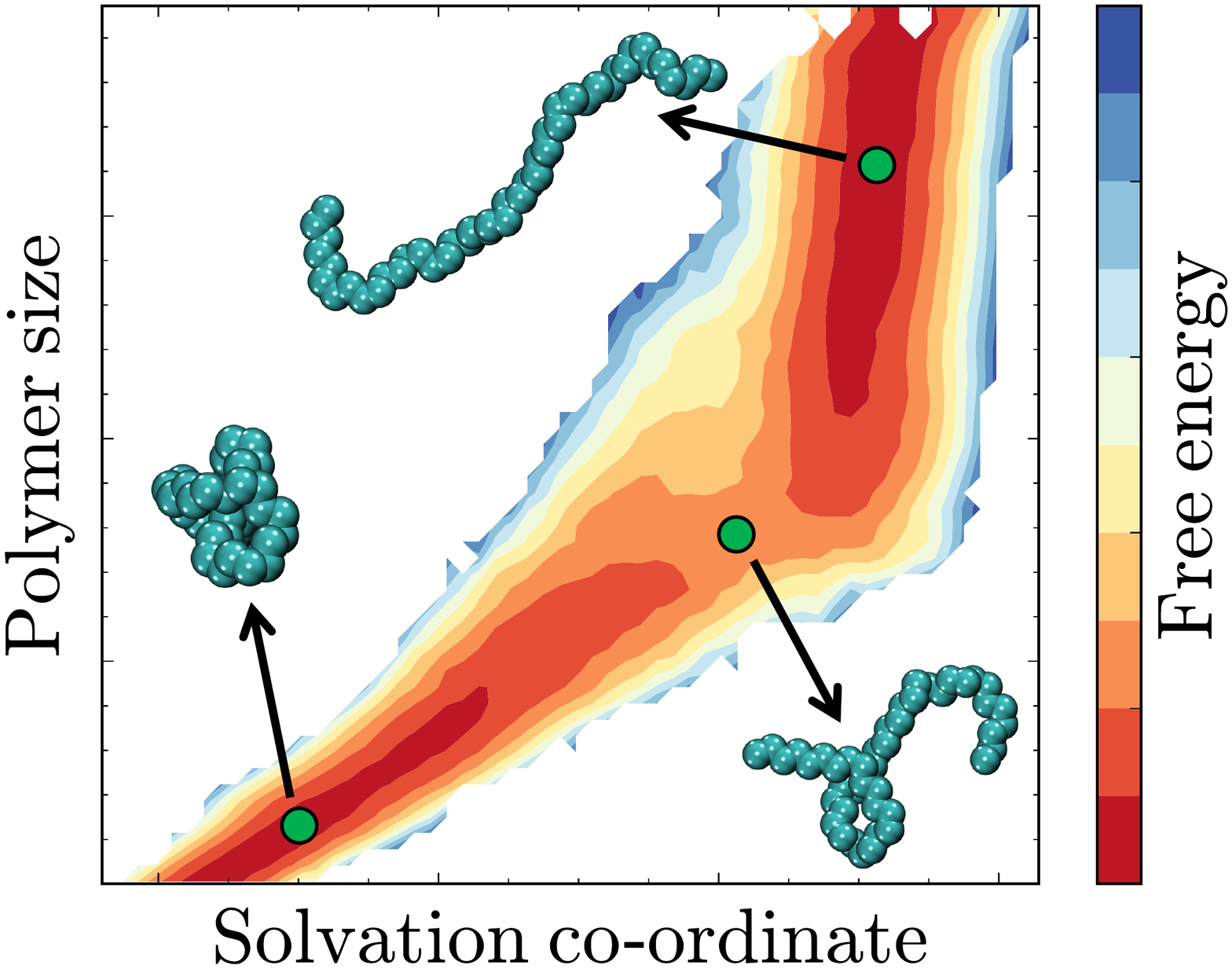}
\end{center}
\end{tocentry}

\oldmaketitle

\begin{abstract}
Conformational transitions of flexible molecules, especially those driven by hydrophobic effects, 
tend to be hindered by desolvation barriers. 
For such transitions, it is thus important to characterize and understand the interplay between solvation and conformation.  
Using specialized molecular simulations, here we perform such a characterization for a hydrophobic polymer 
solvated in water.
We find that an external potential, which unfavorably perturbs the polymer hydration waters, 
can trigger a coil-to-globule or collapse transition, and that the relative stabilities 
of the collapsed and extended states can be quantified by the strength of the requisite potential.
Our results also provide mechanistic insights into the collapse transition, 
highlighting that polymer collapse proceeds through the formation of a sufficiently large non-polar cluster, 
and that collective water density fluctuations play an important role in stabilizing such a cluster.
We also study the collapse of the hydrophobic polymer in octane, a non-polar solvent, 
and interestingly, we find that the mechanistic details of the transition are qualitatively similar to that in water. 
\end{abstract}
\clearpage
%
%
\section{Introduction}
%
Aqueous or organic solutions of conformationally flexible solutes, such as polymers or peptides, 
play an important role in diverse materials~\cite{Rossky:2000:Nature,lin_decreasing_2014,Errington:2015:JPCB,Rick:2016:JPCB} and biomolecular contexts~\cite{thirumalai_chaperonin-mediated_2001,Rossky:2011:JPCB, Shell:2012:JPCB,bellissent2016water}.
Although solvated flexible molecules can exist in a multitude of conformations, they often display two or more distinct basins that are separated by free energy barriers, e.g., polymers can be in extended or collapsed states, 
whereas proteins can be in their native folded structures, become denatured, or even adopt well-defined misfolded configurations~\cite{Geissler:2015:ACR,  Shea:2016:PCCP}.
Molecular simulations have been used extensively, often with the aid of enhanced sampling methods, 
to study a variety of flexible molecules, and characterize their conformational free energy landscapes, 
typically as a function of one or more solute co-ordinates, such as radius of gyration or dihedral angles~\cite{Shell:2012:JPCB,Noid:2014:JPCL}.
%

In characterizing such landscapes, solvent degrees of freedom are usually integrated out. 
Such integration does not lead to any loss of mechanistic information into conformational transitions 
when solvent degrees of freedom equilibrate rapidly with respect to the conformational co-ordinates~\cite{ferguson2010systematic}.
However, certain transitions, such as those driven by hydrophobic effects, tend to 
feature slow solvent degrees of freedom, which relax on timescales that are comparable to, 
or even longer than, the relaxation times of the configurational co-ordinates~\cite{pc02,mvc07,bwr09,Setny1197,rxvsdg15,tiwary2015role}.
In particular, both theoretical and simulation studies have 
highlighted the importance of slow solvent degrees of freedom in the 
coil-to-globule transition of non-polar polymers in water~\cite{lcw99,pc02,mvc07,garde07,garde08}.

For transitions characterized by slow solvent degrees of freedom, 
attempts to integrate out the slow solvent co-ordinates 
can lead to hysteresis in the sampling of the conformational co-ordinates, 
and make it challenging to accurately estimate the conformational free energy landscape~\cite{xmfp18}.
Importantly, the resulting loss of mechanistic information can also obfuscate 
how the solute conformational landscape might respond to changes 
in the solvent, e.g., due to the introduction of co-solutes~\cite{garde08salt,Stirnemann3413,mhlswb15,nv18} or proximity to interfaces~\cite{Jamadagni-2009,la9011839,vpsg13,Shea:2015:Langmuir}.
Thus, characterizing the interplay between conformation and solvation,
e.g., through a free energy landscape that is a function of both conformational and solvent co-ordinates, 
can be valuable, particularly for systems that are expected to feature slow solvent degrees of freedom~\cite{pc02,mvc07,jrrp19}.
However, the enhanced sampling of solvent co-ordinates, which must be performed to obtain such a landscape, 
can be challenging because the solvation shell of a flexible solute is inherently dynamic, 
and changes along with the conformation of the solute. 
%

%
To address this challenge, we recently introduced a method for sampling the number of solvent molecules, $\Ntv$, in a dynamical volume, $v$, which evolves with, and continuously conforms to, the shape of a flexible solute of interest~\cite{jrrp19}. 
Here, we use this method to study the coil-to-globule or collapse transition of a flexible hydrophobic polymer solvated in water. 
We find that polymer collapse can be triggered by an unfavorable potential, $\phi \Ntv$, that perturbs the polymer hydration waters~\cite{pvjhc12,pg14}, and that the potential strength, $\phi$, needed to trigger the transition can serve as a measure of the relative stabilities of the collapsed and extended states.
We also characterize the free energy landscape as a function of both the radius of gyration of the polymer (conformational co-ordinate), and the number of waters, $\Ntv$, in its hydration shell (solvent co-ordinate), 
and use this characterization to uncover mechanistic insights into the collapse transition. 
We find that polymer collapse proceeds through the formation of a sufficiently large non-polar cluster, 
and that collective water density fluctuations play an important role in nucleating such a cluster~\cite{pc02,mvc07}.
We also study the collapse of the hydrophobic polymer in the non-polar solvent, octane, 
and find that it is remarkably similar to the collapse of the polymer in water.

\section{Methods}
To interrogate the interplay between the solvation and conformation of flexible solutes, 
here we study a linear alkane chain consisting of 45 beads that is designated as C$_{45}$.
The solvation shell, $v$, of the C$_{45}$ chain, is defined as the union of 45 spherical sub-volumes with radius, $r_v$, which are pegged to the C$_{45}$ monomers.
Thus, the solvation shell, $v$, of the flexible C$_{45}$ molecule is not static, but changes its shape and/or size dynamically in response to the conformational fluctuations of the polymer.
Using the recently developed dynamic INDUS method~\cite{jrrp19}, 
here we bias the coarse-grained number, $\Ntv$, of solvent heavy atoms in $v$;
$\Ntv$ is closely related to the actual number of solvent atoms in $v$, 
but is chosen to be a continuous function of particle positions to permit biasing $\Ntv$ 
without resulting in impulsive forces~\cite{pvc10,pvc11}.
By using harmonic biasing potentials to sample $\Ntv$ over its range of interest, 
and using the Weighted Histogram Analysis Method (WHAM) 
to combine the biased probability distributions~\cite{roux,mbar,zh12,UWHAM}, 
we are able to characterize the free energetics, $\beta \gvnt \equiv -\ln \pvnt$, 
where $\beta^{-1} \equiv \kbt$ is the thermal energy, and $\pvnt$ 
is the probability of observing $\Nt$ solvent atoms in $v$.
The probability, $\pvnt \equiv \langle \delta(\Nt - \Ntv) \rangle_0$, is the ensemble average
of the Dirac delta function obtained using the unbiased Hamiltonian, $\mathcal{H}_0$.
Once $\gvnt$ is estimated, we use standard reweighting techniques~\cite{roux} 
to interrogate how biasing potentials, such as $U_\phi \equiv \phi \Ntv$, 
modulate polymer solvation and determine the corresponding averages, $\langle \dots \rangle_\phi$, 
obtained using the biased Hamiltonians, $\mathcal{H}_\phi = \mathcal{H}_0 + U_\phi$.
For example, the free energetics in a biased ensemble, $\beta G_v^\phi(\Nt) \equiv -\ln P_v^{\phi}(\Nt)$, 
can be obtained using $P_v^{\phi}(\Nt) = \langle \delta(\Nt - \Ntv) \rangle_\phi$.
Finally, to characterize the two-dimensional free energetics, $\beta G_v(\Nt,\Rg) \equiv -\ln P_v(\Nt,\Rg)$, 
where $P_v(\Nt,\Rg)$ is the joint probability of observing $\Nt$ solvent atoms in $v$, and the polymer with a radius of gyration, $\Rg$, 
we combine the joint probability distributions, observed in our biased simulations, 
using the WHAM weights determined above~\cite{pvc11,xrp16}.

\subsection{Simulation Details}
\label{sec:comp_det}
All simulations were performed using GROMACS package~\cite{hkvl08} (version 4.5.3), suitably modified to 
perform indirect umbrella sampling (INDUS) using dynamic probe volumes~\cite{pvc11,jrrp19}. 
The equations of motion were integrated using the leap-frog algorithm with a time step of 2~fs, and periodic boundary conditions were employed in all three dimensions. 
The system temperature, $T$, was maintained at 298~K using the canonical velocity rescaling thermostat~\cite{Bussi:JCP:2007} with a time constant of 0.5 ps, and the system pressure, $P$, was maintained at 1~atm using the Parrinello-Rahman barostat~\cite{Parrinello-Rahman} with a time constant of 1~ps. 
The n-alkane chain containing 45 carbons (designated as C$_{45}$) was modeled using the TraPPE-UA (Transferable Potentials for Phase Equilibria -- United Atom) forcefield developed by Siepmann~{\it et.~al.}~\cite{ms98}.
The TraPPE-UA forcefield models individual -CH$_2$- or -CH$_3$ units as pseudo-atoms, so that our solute is comprised of 45 united-atom beads.
The C$_{45}$ chain was solvated using either water or n-octane, and Lorentz-Berthelot combination rules 
were used to determine the Lennard Jones parameters for cross-interactions.
In all cases, the simulation box was first energy minimized using the steepest descent algorithm, 
followed by a 2~ns NVT simulation, and then a 4~ns NPT simulation to equilibrate the system.
The biased simulations were run for 10~ns, with either the first 2~ns (for C$_{45}$/water) or 1~ns (for C$_{45}$/octane) being discarded for equilibration; the coarse-grained number of solvent atoms, $\Ntv$, in the polymer hydration shell, $v$, and the polymer radius of gyration, $\Rg$, were estimated every 0.1~ps. 
Additional details of the  C$_{45}$/water and C$_{45}$/octane simulations are included below. 

\noindent{\bf C$_{45}$/Water: } 
Water was modeled using the SPC/E (extended simple point charge) model~\cite{bgs87} with bonds involving hydrogen atoms being constrained using the SETTLE algorithm~\cite{SETTLE}.
The particle Mesh Ewald (PME) method~\cite{PME} was used to compute long-range electrostatic interactions; 
short-range electrostatic and Lennard Jones interactions were truncated at 1~nm. 
The C$_{45}$ chain was hydrated using 11,614 water molecules giving rise to a roughly $7$~nm cubic simulation box.
Before performing biased simulations, a 10~ns long unbiased simulation was performed to estimate the average number of solvent atoms in $v$, and ascertain the $\Ntv$-range that must be sampled;
the radius of the spherical sub-volumes used to define $v$ was chosen to be $r_v = 0.6$~nm. 
The biased simulations used harmonic potentials with a spring constant of 0.0243~kJ/mol. 
%

\noindent{\bf C$_{45}$/Octane:} 
Octane was modeled using the TraPPE-UA forcefield~\cite{ms98} with Lennard Jones interactions being truncated at 1.4~nm.
The C$_{45}$ chain was solvated using 1,000 octane molecules.
The sub-volume radius used to define $v$ was chosen to be $r_v = 0.7$~nm,
and a spring constant of 0.033~kJ/mol was used for the harmonic biasing potentials. 
%

\section{Results and Discussion}
\subsection{How Polymer Hydration Waters Respond to Perturbations}
\begin{figure*}[t!]
  \centering
  \includegraphics[clip=true,width=16.5cm]{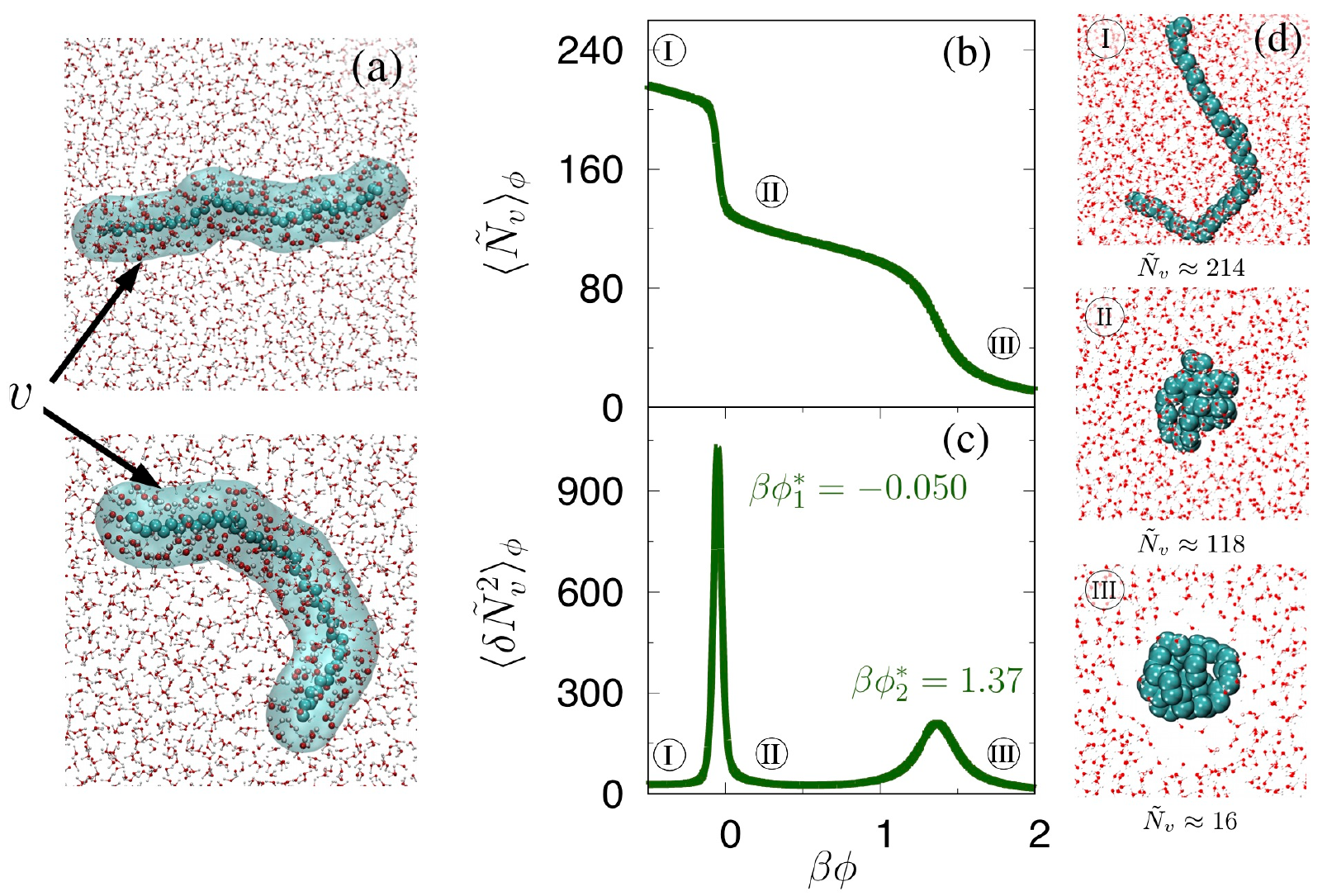}
\caption{
(a) Simulation snapshots of the \cforty~polymer (cyan, space-fill) solvated in water (red/white) are shown. 
Water molecules in the polymer hydration shell, $v$ (cyan, transparent), are highlighted.
As the polymer undergoes conformational fluctuations, both the shape and size of its hydration shell change; 
to account for such changes, $v$ is defined to be the union of 45 spherical sub-volumes 
that are centered on the polymer beads and have radii, $r_v = 0.6$~nm.
(b) As the strength, $\phi$, of a linear biasing potential, $\Uphi = \phi \Ntv$, is increased,
the average number of waters, $\avgNtphi$, in the hydration shell decreases; 
interestingly, two sharp drops are observed in $\avgNtphi$.
(c) Correspondingly, the susceptibility, $-\partial \avgNtphi / \partial(\beta \phi) = \varNtphi$, 
displays two peaks at $\phi$-values denoted by $\phi_1^*$ and $\phi_2^*$.
(d)~Simulation snapshots corresponding to three select $\Ntv$-values are shown.
As shown in snapshot~\one, the polymer adopts an extended configuration for $\Ntv > \avgNt_{\phi_1^*}$,
whereas snapshot~\two highlights that the polymer is in a collapsed configuration for $\avgNt_{\phi_2^*} < \Ntv < \avgNt_{\phi_1^*}$.
As illustrated in snapshot~\three, the collapsed polymer is dewetted for $\Ntv < \avgNt_{\phi_2^*}$.
} 
\label{fig1}
\end{figure*}

To characterize how the hydration of the flexible C$_{45}$ chain influences its conformation, 
we focus on water molecules in its first hydration shell, $v$ (Figure~\ref{fig1}a, cyan/transparent). 
We account for the dynamic nature of the polymer hydration shell, i.e., changes in its 
size and shape in response to polymer conformational fluctuations,
by defining $v$ to be the union of 45 spherical sub-volumes, 
where every sub-volume is pegged to a polymer bead, and has a radius, $r_v = 0.6$~nm.
We then interrogate how the polymer hydration waters respond to perturbations 
by employing a biasing potential, $\Uphi = \phi \Ntv$, where $\Ntv$ is the 
(coarse-grained) number of waters in $v$, and $\phi$ is the potential strength. 
Such a biasing potential provides a convenient way to modulate $\Ntv$, and thereby polymer hydration.
In particular, hydration is disfavored for $\phi > 0$ with the perturbation becoming 
more pronounced as $\phi$ is increased; conversely, for $\phi < 0$, polymer hydration is favored.
From a physical standpoint, a biasing potential of strength, $\phi$, effectively decreases the pressure in the polymer hydration shell by roughly $\phi \rho_{\rm w}$, where $\rho_{\rm w}$ is the density of bulk water~\cite{rxp19}.

The response of the C$_{45}$ hydration waters to the biasing potential is shown in Figure~\ref{fig1}.
As expected, the average number of hydration waters, $\avgNtphi$, decreases as $\phi$ is increased (Figure~\ref{fig1}b).
However, the decrease in $\avgNtphi$ is not gradual, but is punctuated by two sharp drops.
The corresponding susceptibility, $\partial \avgNtphi / \partial(-\beta \phi) = \varNtphi$, shown in Figure~\ref{fig1}c, displays marked peaks at $\phi$-values, which are denoted by $\phi_1^*$ and $\phi_2^*$ .
The sharp drops in $\avgNtphi$, and the corresponding peaks in susceptibility, suggest 
that the polymer undergoes collective transitions at potential strengths, $\phi_1^*$ and $\phi_2^*$.
To shed light on the nature of these transitions, we include representative snapshots of the system 
for select $\Ntv$-values in Figure~\ref{fig1}d.
Snapshot~\one suggests that the polymer is in an extended configuration for $\phi < \phi_1^*$;
snapshot~\two shows the polymer in a collapsed, but hydrated configuration for $\phi_1^* < \phi < \phi_2^*$;
and snapshot~\three highlights that the collapsed polymer is dewetted for $\phi > \phi_2^*$.
These snapshots suggest that at potential strength, $\phi_1^*$, the polymer undergoes a collapse (or folding) transition, whereas at potential strength, $\phi_2^*$, the collapsed polymer dewets.
%

\begin{figure*}[h!]
  \centering
  \includegraphics[clip=true,width=16.5cm]{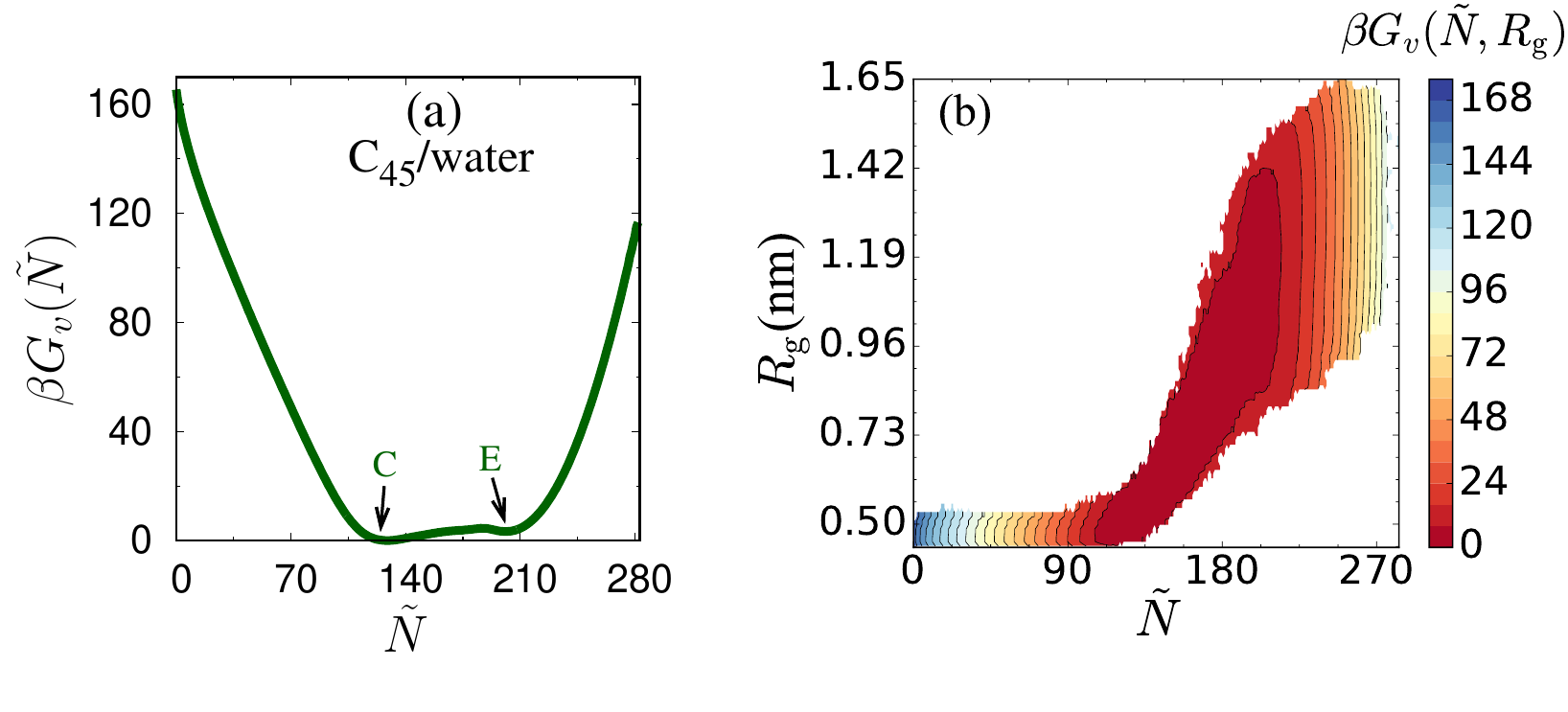}
\caption{
(a) The free energetics of water number fluctuations, $G_v(\Nt)$, in the dynamical hydration shell of the \cforty~polymer are shown. 
The basins corresponding to the collapsed (C) and extended (E) states of the polymer are denoted with arrows. 
(b) To illustrate the interplay between the hydration of the hydrophobic polymer in water and its conformation,
we plot the two-dimensional free energy landscape, $G_v(\Nt,\Rg)$, where $\Nt$ is the number of waters in $v$, 
and $\Rg$ is the radius of gyration of the \cforty~polymer. 
The relatively sharp increase in $G_v(\Nt, \Rg)$ at low $\Nt$ and $\Rg$ corresponds to 
the free energetic cost of dewetting the collapsed polymer globule; 
whereas the increase in $G_v(\Nt, \Rg)$ at high $\Nt$ and $\Rg$ corresponds to the 
unfavorable wetting of the extended polymer.
}
\label{fig2}
\end{figure*}

\subsection{Interplay Between Polymer Hydration and Conformation} 
To characterize the free energy landscape underpinning the collapse and dewetting transitions, 
we estimate the free energetics, $\gvnt$, of water number fluctuations in the hydration shell 
of the C$_{45}$ polymer (Figure~\ref{fig2}a), as well as the corresponding 
two-dimensional free energy landscape, $G_v(\Nt, \Rg)$ (Figure~\ref{fig2}b).
The conformational free energy landscape, $G_v(\Rg)$, can be readily obtained from $G_v(\Nt, \Rg)$ 
by integrating out $\Nt$, and is included in Figure~S\ref{fig1} of the Supplementary Material. 
Figure~\ref{fig2} confirms the presence of two nearly degenerate basins, 
observed at low $\Nt \approx 130$ and $\Rg \approx 0.5$~nm, 
and at higher $\Nt \approx 200$ and $\Rg \approx 1.2$~nm, 
corresponding to the collapsed (C) and extended (E) states, respectively.
Figure~\ref{fig2} also suggests that the solvation and conformational co-ordinates 
are coupled to one another across the collapse transition, and that 
the transition incurs a free energetic cost of less than $8~\kbt$.
In contrast, decreasing $\Nt$ below 100 results in a sharp increase in $G_v(\Nt,\Rg)$; 
such a decrease in $\Nt$ is not accompanied by a further decrease in $\Rg$, 
suggesting that it corresponds to the dewetting of the collapsed polymer.
Similarly, increasing $\Nt$ above 220 also incurs a large free energetic penalty, 
and is not accompanied by an increase in $\Rg$, 
suggesting that it represents the unfavorable compression of 
waters in the hydration shell of the extended polymer.
%

%
Although the free energy difference between the collapsed and extended states 
or the folding free energy, $\Delta G_{\rm fold}$ is relatively small, 
Figure~\ref{fig2}a suggests that for the hydrophobic $C_{45}$ chain in water, 
the collapsed state is nevertheless stable relative to the extended state, 
with $\beta\Delta G_{\rm fold} = -3.3$.
The relative stability of the collapsed state is also evident from the fact that the potential strength, $\phi^*_1$, corresponding to the collapse transition is negative (Figure~\ref{fig1}c).
In fact, as shown below, the two quantities are proportional to one another, with $\Delta G_{\rm fold} \approx \phi_1^*(\Nte - \Ntc)$, where $\Nte$ and $\Ntc$ are $\Nt$-values corresponding to the extended and collapsed basins, respectively.
The collapsed and extended states are expected to be in coexistence with one another 
in the $\phi^*_1$-ensemble, i.e., $G_{v}^{\phi_1^*}(\Ntc) = G_{v}^{\phi_1^*}(\Nte)$, and
because $G_{v}^{\phi}(\Nt)$ equals $G_{v}(\Nt) + \phi\Nt$ within a constant~\cite{xrp16}:
$G_{v}(\Ntc) + \phi_1^* \Ntc = G_{v}(\Nte) + \phi_1^* \Nte$;
thus, the polymer folding free energy, $\Delta G_{\rm fold} \approx G_{v}(\Ntc) - G_{v}(\Nte) = \phi_1^*(\Nte - \Ntc)$.
For the C$_{45}$ polymer in water, we estimate $\phi_1^*(\Nte - \Ntc) \approx -3.47~\kbt$, which in good agreement with $\Delta G_{\rm fold}$ obtained directly from $\gvnt$.
Thus, $\phi^*_1$ can serve as a quantitative measure of $\Delta G_{\rm fold}$.

\subsection{Free Energy Landscape at the Collapse Transition}
\begin{figure*}[h!]
  \centering
  \includegraphics[clip=true,width=16.5cm]{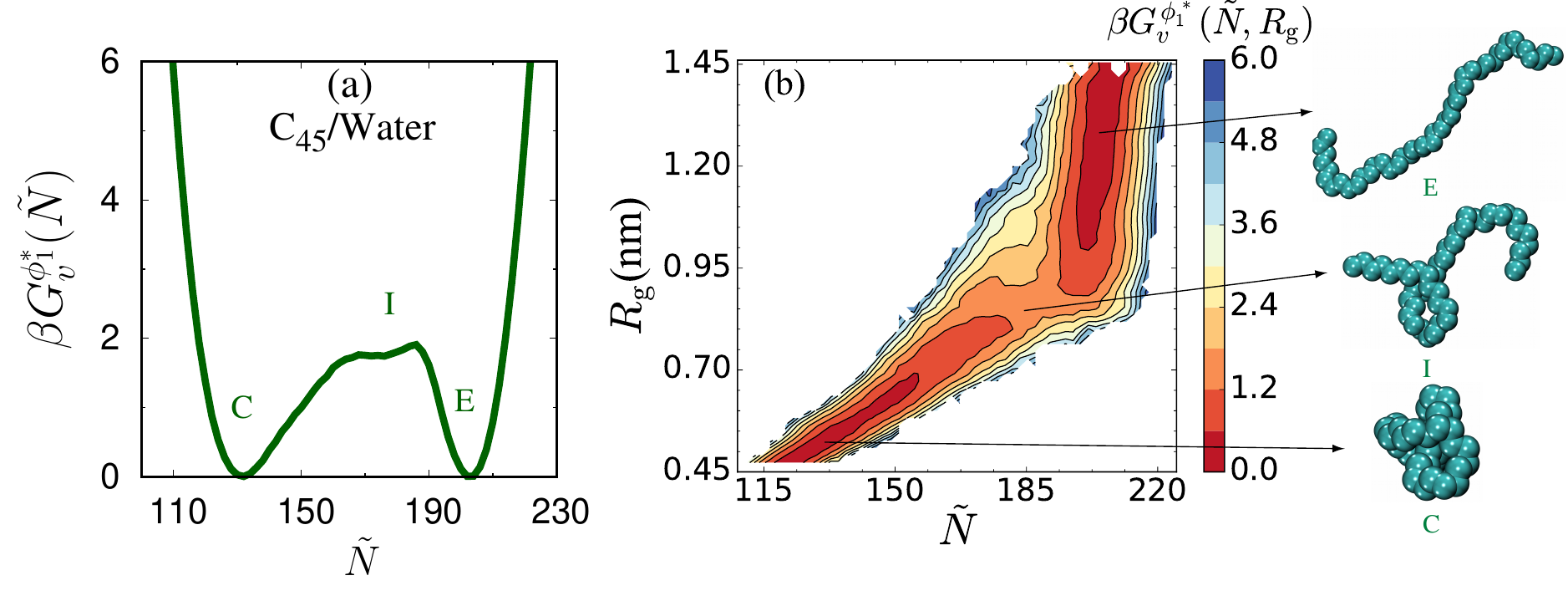}
\caption{
To better understand the coil-to-globule collapse transition of the \cforty~polymer, 
here we study its behavior in the $\phi_1^*$-ensemble; 
that is, in the presence of the biasing potential, $\phi_1^* \Ntv$, where $\beta\phi_1^* = -0.05$.
In such a biased ensemble, the coil and globule states are expected to be in coexistence 
with one another. 
(a) The free energetics, $G_v^{\phi_1^*}(\Nt)$, of water number fluctuations in the $\phi_1^*$-ensemble,  
display two distinct basins that are separated by a barrier of roughly $2~\kbt$.
The basins corresponding to the collapsed (low $\Nt$) and extended (high $\Nt$) states of the polymer 
are labelled ``C'' and ``E'', respectively, whereas the location of the maximum in $G_v^{\phi_1^*}(\Nt)$ 
at intermediate $\Nt$ is denoted by ``I''. 
(b) The two-dimensional free energy landscape in the $\phi^*_1$-ensemble, $G_v^{\phi_1^*}(\Nt,\Rg)$, is shown here, 
and sheds light on the minimum free energy path that the polymer is likely to adopt as it undergoes collapse.
Representative configurations of the polymer in the ``C'', ``I'', and ``E'' states are also shown.
} 
\label{fig3}
\end{figure*}

To better understand the transition of the C$_{45}$ polymer between its extended and collapsed states, 
we now focus on its behavior in the presence of a biasing potential of strength, $\phi_1^*$, 
wherein the two states are expected to be in coexistence with one another. 
The corresponding free energy landscapes, $G_v^{\phi^*_1}(\Nt)$ and $G_v^{\phi^*_1}(\Nt,\Rg)$ are shown in Figure~\ref{fig3}.
As seen in Figure~\ref{fig3}a, $G_v^{\phi^*_1}(\Nt)$ features two basins: 
the low-$\Nt$, collapsed (C) and the high-$\Nt$, extended (E) basins, 
which are in coexistence with one another, and are separated by a barrier of roughly $2~\kbt$.
Figure~\ref{fig3}a also highlights the locations of the basins (C, E) and the maximum in $G_v^{\phi^*_1}(\Nt)$ at intermediate $\Nt$ (I). 

%
The two-dimensional free energy landscape, $G_v^{\phi^*_1}(\Nt,\Rg)$, shown in Figure~\ref{fig3}b,
sheds light on the mechanistic pathways that the polymer follows as it undergoes the collapse transition, 
i.e., it elucidates the minimum free energy path in the ($\Nt$, $\Rg$) space, 
with $\Nt$ and $\Rg$ representing the polymer hydration and conformational degrees of freedom.
The $G_v^{\phi^*_1}(\Nt,\Rg)$ landscape highlights that in its extended state, 
the polymer can undergo substantial fluctuations in $\Rg$, but only small fluctuations in $\Nt$.
To approach the saddle point in $G_v^{\phi^*_1}(\Nt,\Rg)$,
the polymer must undergo a marked low-$\Rg$ fluctuation, 
which is followed by a substantive decrease in its hydration waters, $\Nt$. 
Upon traversing the saddle point, polymer collapse proceeds through a 
co-ordinated decrease in both $\Rg$ and $\Nt$, 
which is downhill in free energy.

Representative configurations of the polymer in the extended (E), saddle point (I) and collapsed (C) states 
are also shown Figure~\ref{fig3}b, and shed further light on the pathways involved in polymer collapse.
In particular, at the saddle point, the polymer adopts a partially collapsed conformation 
with some, but not all, of its monomers being clustered together.
Thus, the formation of a sufficiently large hydrophobic cluster, which not only lowers the $\Rg$ of the polymer, 
but also leads to a decrease in its hydration waters, represents the barrier to polymer collapse.
Importantly, as the polymer crosses this barrier, the decrease in its $\Rg$ is small 
relative to the corresponding decrease in its hydration waters, $\Nt$.
These findings highlight the importance of collective water density fluctuations in facilitating polymer collapse.

Our results also lend support to the observations of ten Wolde and Chandler~\cite{pc02}, 
who studied the collapse of a purely repulsive, idealized hydrophobic polymer using a coarse-grained model of water, and those of Miller {\it et al.}\cite{mvc07}, who studied the same polymer using atomistic simulations; 
both studies found that the formation of a critical non-polar cluster 
and the associated dewetting represent the barrier to polymer collapse.
Moreover, the agreement between these studies, performed on ideal hydrophobic polymers, 
and our results, obtained using realistic polymers, which have favorable dispersion interactions with water, 
highlight that weak polymer-water attractions do not qualitatively alter the polymer hydration landscape~\cite{rp15}.

\subsection{Polymer Collapse in Octane}
\begin{figure*}[h!]
  \centering
  \includegraphics[width=\textwidth]{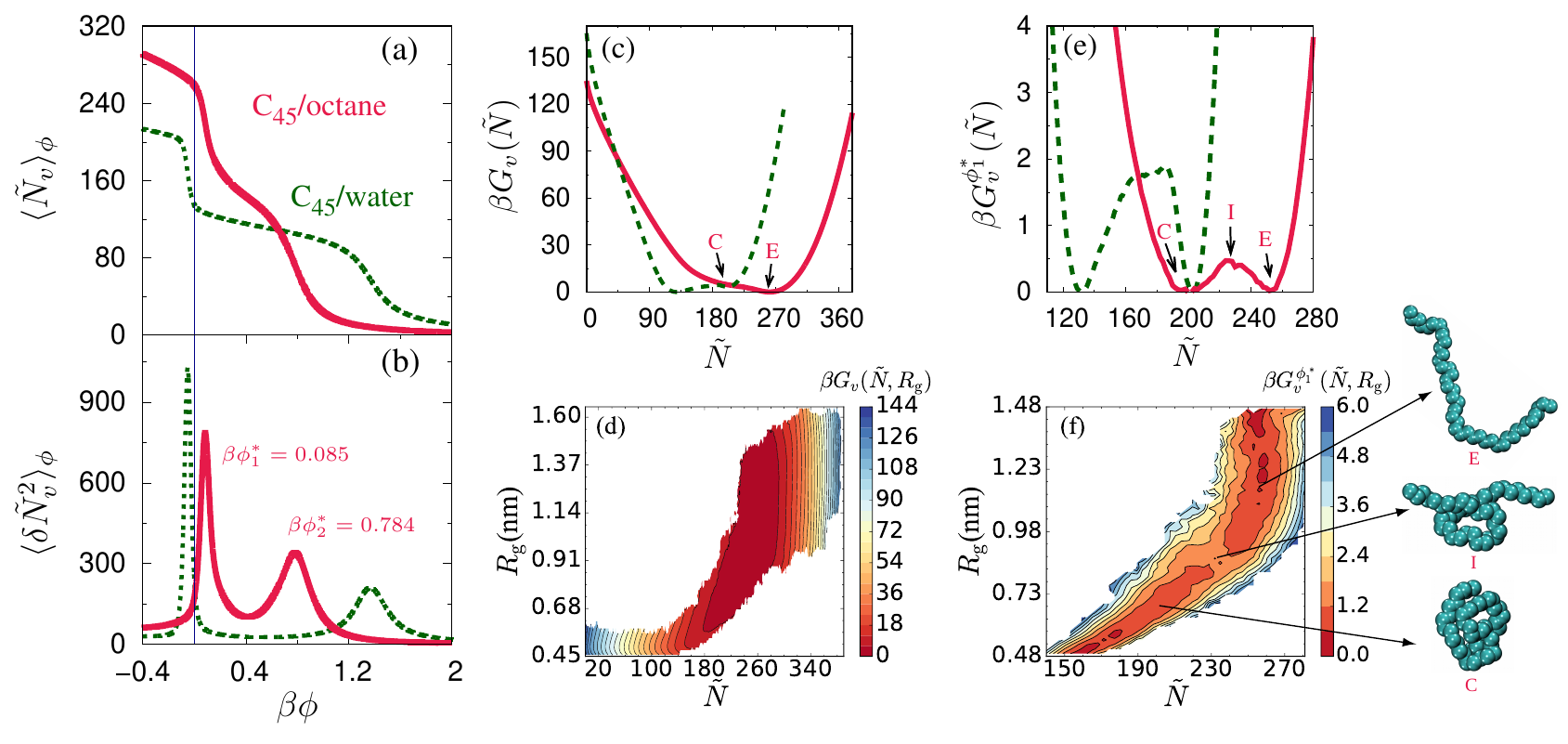}
\caption{
The \cforty~polymer solvated in octane. 
(a) The average number of solvent heavy atoms, $\avgNtphi$, in the solvation shell, $v$, of the \cforty~polymer
are shown as a function of the strength, $\phi$, of the linear biasing potential, $\Uphi = \phi \Ntv$.
For the \cforty~polymer solvated in both octane (red, solid) and in water (green, dashed lines), two sharp drops in $\avgNtphi$ are observed as $\phi$ is increased; the vertical line (blue) depicts $\phi = 0$. 
(b) Correspondingly, the susceptibility, $-\partial \avgNtphi / \partial(\beta \phi) = \varNtphi$, 
displays two peaks in both cases; the locations of the peaks, $\phi_1^*$ and $\phi_2^*$, are included for octane.
In contrast with \cforty~in water, $\phi_1^* > 0$ for \cforty~solvated in octane, 
suggesting that polymer is in its extended state at equilibrium ($\phi =0$).
(c) The free energetics of solvent number fluctuations, $G_v(\Nt)$, in the hydration shell 
of the \cforty~polymer are shown, and the locations of the collapsed (C) and extended (E) states 
of the polymer are denoted with arrows. 
In octane, the extended state (high $\Nt$) is stable, whereas the collapsed state (low $\Nt$) is unstable.
(d) The two-dimensional free energy landscape, $G_v(\Nt,\Rg)$, for the \cforty~polymer solvated in octane is shown. %
(e) The free energetics, $G_v^{\phi_1^*}(\Nt)$, of solvent number fluctuations in the $\phi_1^*$-ensemble,  
display two distinct basins that are separated by a barrier. 
(f) The two-dimensional free energy landscape, $G_v^{\phi_1^*}(\Nt,\Rg)$, 
shown here for the \cforty~polymer solvated in octane, 
as well as the representative configurations of the polymer, 
are qualitatively similar to those for \cforty~in water (Figure~\ref{fig3}b).
} 
\label{fig4}
\end{figure*}

We now study the solvation of the hydrophobic C$_{45}$ polymer in octane, a non-polar solvent, 
and draw comparisons to the solvation of the polymer in water.
In Figure~\ref{fig4}a, we plot the response of the average number of solvent heavy atoms, $\avgNtphi$, 
 in the polymer solvation shell, $v$, to the strength, $\phi$ of the linear biasing potential, $U_\phi = \phi \Ntv$, 
 and in Figure~\ref{fig4}b, we plot the corresponding susceptibility, $-\partial\avgNtphi/\partial(\beta\phi) = \varNtphi$. 
Once again, two sharp drops are observed in $\avgNtphi$ with increasing $\phi$, and correspondingly, 
two peaks are seen in $\varNtphi$ at $\phi$-values denoted by $\phi_1^*$ and $\phi_2^*$.
The configurations of C$_{45}$ in octane, shown for three select $\Ntv$-values in Figure~S\ref{fig2} of the Supplementary Material, suggest that the transitions at $\phi_1^*$ and $\phi_2^*$, once again, correspond to polymer collapse and the dewetting of the collapsed polymer. 
Vertical lines in Figures~\ref{fig4}a,b (blue) correspond to $\phi = 0$, and
highlight that in contrast with C$_{45}$ in water, $\phi_1^*$ is positive for C$_{45}$ in octane.
Thus, the folding (or collapse) free energy, $\Delta G_{\rm fold} \approx \phi^*_1(\Nte - \Ntc)$, must also be positive,
suggesting that in octane, the extended state of C$_{45}$ is stable relative to its collapsed state. 
Figure~\ref{fig4}b also shows that the dewetting of the collapsed polymer occurs at a lower $\phi^*_2$-value 
in octane than in water; this observation can be readily rationalized in terms of the smaller 
liquid-vapor surface tension of octane, which makes it easier to form cavities~\cite{wg15}.
%

%
The unbiased free energy landscapes, $G_v(\Nt)$ and $G_v(\Nt, \Rg)$, for C$_{45}$ in octane, are shown in Figures~\ref{fig4}c,d, and the conformational free energy landscape, $G_v(\Rg)$, is included as Figure~S\ref{fig3} of the Supplementary Material.
Apart from the fact that the collapsed state is metastable with respect to the extended state, with $\beta\Delta G_{\rm fold} = 4.5$, the landscapes are qualitatively similar to those for C$_{45}$ in water (Figure~\ref{fig2}).
The corresponding free energy landscapes in the $\phi_1^*$-ensemble are shown in Figures~\ref{fig4}e,f. 
As with C$_{45}$ in water, the collapsed (C) and extended (E) states are degenerate in the $\phi_1^*$-ensemble, 
and a barrier is observed at intermediate $\Nt$ and $\Rg$ for C$_{45}$ in octane.
Although the barrier height is somewhat lower for C$_{45}$ in octane (relative to water), 
both the $G_v^{\phi_1^*}(\Nt, \Rg)$ landscape and representative configurations of the polymer (Figure~\ref{fig4}f), 
suggest that polymer collapse in octane proceeds through mechanistic pathways, 
which are remarkably similar to those in water (Figure~\ref{fig3}b).
In particular, as in water, the collapse of the C$_{45}$ polymer in octane 
also proceeds through the formation of a sufficiently large cluster, 
which is both partially collapsed and partially desolvated. 

%
To understand these similarities, we first recognize that polymer collapse involves 
the solvent-mediated assembly of well-solvated sub-units, which are smaller than 1~nm, 
into a cluster that is larger than 1~nm in size~\cite{Li-Walker,gp11}.
According to the Lum-Chandler-Weeks theory of hydrophobicity~\cite{lcw99}, 
the solvation free energy of the former scales as their excluded volume, 
whereas the solvation free energy of the latter scales as its surface area~\cite{hggpp96,rts05,vpc11,Geissler:2016:PNAS,xp16}; this interplay between solvation at small and large length scales
is responsible for the emergence of a critical cluster, which must be nucleated for the collapse transition to proceed~\cite{pc02,mvc07}.
Interestingly, a solute size-dependent crossover in solvation free energies has also been observed in non-polar solvents, such as octane~\cite{Huang:2000ve,wg15}.
Although the corresponding solvophobic effect, experienced by solutes in non-polar solvents, 
is expected to be weaker than the hydrophobic effect~\cite{rts05,pgd:JPCL:2011},
our results suggest that it may nevertheless be responsible for the qualitative similarities 
between the polymer collapse pathways in water and in octane.

\section{Conclusions and Outlook}
Using the recently developed dynamic Indirect Umbrella Sampling (INDUS) method for sampling the number of solvent molecules, $\Ntv$, in the dynamical solvation shell of a flexible solute, here we study the 
coil-to-globule or collapse transition of a hydrophobic polymer solvated in water and in the non-polar solvent, octane.
We find that an unfavorable potential, which perturbs $\Ntv$, can trigger the collapse transition, and that the requisite potential strength quantifies the relative stabilities of the collapsed and extended states.
To understand the interplay between polymer hydration and conformation, 
we characterize the free energy landscape of the system as a function of both the number of 
solvent molecules, $\Ntv$, in the polymer hydration shell, and its radius of gyration, $\Rg$. 
%
We find that polymer collapse proceeds through the formation of a critical non-polar cluster, 
which represents the bottleneck to the coil-to-globule transition.
In particular, assisted by thermal fluctuations, an extended polymer must first 
adopt a relatively compact configuration, which then undergoes collective dewetting; 
if the resulting non-polar cluster is sufficiently large, the rest of the polymer then collapses spontaneously.

In agreement with previous studies on ideal hydrophobic polymers~\cite{pc02,mvc07}, 
our findings thus confirm the presence of slow solvent co-ordinates in the coil-to-globule transition, 
and highlight the importance of collective solvent density fluctuations in overcoming the corresponding barriers.
In contrast with water, which is a poor solvent for the hydrophobic polymer, the non-polar solvent, octane, is expected to be a good solvent for the polymer. 
Interestingly, we find that the mechanistic details of the collapse transition in octane are nevertheless remarkably similar to that in water.
Our results suggest that the size-dependent crossover in solvation free energies, 
which has been observed not just in water~\cite{rts05}, but also in non-polar solvents, 
such as octane~\cite{Huang:2000ve,wg15},
may underpin the similar polymer collapse pathways in these two solvents.
%

A characterization of the free energetics of water density fluctuations in static volumes, 
both in bulk water~\cite{ghgpp96,hggpp96} and in the vicinity of surfaces~\cite{gjg09,Acharya:Faraday:2010,Rotenberg:JACS:2011,Jamadagni:ARCB:2011},
has provided numerous insights into the collective, solvent-mediated hydrophobic effects, 
ranging from micelle formation and interfacial assembly, to superhydrophobicity and protein interactions~\cite{micelle,pvjagc11,pxp16,xvlrpg14,Rego:2021gx}.
By characterizing the free energetics of solvent density fluctuations 
in the dynamic solvation shells of conformationally flexible molecules, 
we hope that our work will similarly pave the way for a richer understanding of 
the conformational transitions~\cite{rhv15,yanm15} and the phase behavior~\cite{dzbkm18,bv19} of a large classes of interesting solutes, 
ranging from hydrocarbons and polymers, to peptides and nucleic acids~\cite{Shea:2015:Langmuir,elise-dna,jrrp19}.
Our approach is also likely to shed light into how co-solutes (e.g., osmolytes, salts, etc.), 
which have a propensity to be included or excluded from the vicinity of a polymer, 
might modulate the solvation of a polymer as well as its conformational landscape~\cite{Shea:2016:JPCL,nv17,Remsing:JPCB:2018,cremer-arpc,okur2017beyond,mwmswm19}.

\begin{acknowledgement}
A.J.P. gratefully acknowledges financial support from the National Science Foundation (NSF grants CBET-1652646, CHE-1665339 and DMR-1720530), and awards from the Alfred P. Sloan Research Foundation (FG-2017-9406) 
and the Camille and Henry Dreyfus Foundation (TC-19-033). 
D.D. was supported by NSF grant CHE-1665339. 
The authors thank Akash Pallath for helpful discussions.
\end{acknowledgement}



\begin{mcitethebibliography}{81}
\providecommand*\natexlab[1]{#1}
\providecommand*\mciteSetBstSublistMode[1]{}
\providecommand*\mciteSetBstMaxWidthForm[2]{}
\providecommand*\mciteBstWouldAddEndPuncttrue
  {\def\EndOfBibitem{\unskip.}}
\providecommand*\mciteBstWouldAddEndPunctfalse
  {\let\EndOfBibitem\relax}
\providecommand*\mciteSetBstMidEndSepPunct[3]{}
\providecommand*\mciteSetBstSublistLabelBeginEnd[3]{}
\providecommand*\EndOfBibitem{}
\mciteSetBstSublistMode{f}
\mciteSetBstMaxWidthForm{subitem}{(\alph{mcitesubitemcount})}
\mciteSetBstSublistLabelBeginEnd
  {\mcitemaxwidthsubitemform\space}
  {\relax}
  {\relax}

\bibitem[Hu \latin{et~al.}(2000)Hu, Yu, Wong, Bagchi, Rossky, and
  Barbara]{Rossky:2000:Nature}
Hu,~D.; Yu,~J.; Wong,~K.; Bagchi,~B.; Rossky,~P.~J.; Barbara,~P.~F. Collapse of
  stiff conjugated polymers with chemical defects into ordered, cylindrical
  conformations. \emph{Nature} \textbf{2000}, \emph{405}, 1030--1033\relax
\mciteBstWouldAddEndPuncttrue
\mciteSetBstMidEndSepPunct{\mcitedefaultmidpunct}
{\mcitedefaultendpunct}{\mcitedefaultseppunct}\relax
\EndOfBibitem
\bibitem[Lin \latin{et~al.}(2014)Lin, Martin, and
  Jayaraman]{lin_decreasing_2014}
Lin,~B.; Martin,~T.~B.; Jayaraman,~A. Decreasing {Polymer} {Flexibility}
  {Improves} {Wetting} and {Dispersion} of {Polymer}-{Grafted} {Particles} in a
  {Chemically} {Identical} {Polymer} {Matrix}. \emph{ACS Macro Letters}
  \textbf{2014}, \emph{3}, 628--632\relax
\mciteBstWouldAddEndPuncttrue
\mciteSetBstMidEndSepPunct{\mcitedefaultmidpunct}
{\mcitedefaultendpunct}{\mcitedefaultseppunct}\relax
\EndOfBibitem
\bibitem[Pandav \latin{et~al.}(2015)Pandav, Pryamitsyn, Errington, and
  Ganesan]{Errington:2015:JPCB}
Pandav,~G.; Pryamitsyn,~V.; Errington,~J.; Ganesan,~V. Multibody Interactions,
  Phase Behavior, and Clustering in Nanoparticle--Polyelectrolyte Mixtures.
  \emph{J. Phys. Chem. B.} \textbf{2015}, \emph{119}, 14536--14550\relax
\mciteBstWouldAddEndPuncttrue
\mciteSetBstMidEndSepPunct{\mcitedefaultmidpunct}
{\mcitedefaultendpunct}{\mcitedefaultseppunct}\relax
\EndOfBibitem
\bibitem[Sharma \latin{et~al.}(2016)Sharma, Liu, Parameswaran, Grayson,
  Ashbaugh, and Rick]{Rick:2016:JPCB}
Sharma,~A.; Liu,~L.; Parameswaran,~S.; Grayson,~S.~M.; Ashbaugh,~H.~S.;
  Rick,~S.~W. Design of Amphiphilic Polymers via Molecular Dynamics
  Simulations. \emph{J. Phys. Chem. B} \textbf{2016}, \emph{120},
  10603--10610\relax
\mciteBstWouldAddEndPuncttrue
\mciteSetBstMidEndSepPunct{\mcitedefaultmidpunct}
{\mcitedefaultendpunct}{\mcitedefaultseppunct}\relax
\EndOfBibitem
\bibitem[Thirumalai and Lorimer(2001)Thirumalai, and
  Lorimer]{thirumalai_chaperonin-mediated_2001}
Thirumalai,~D.; Lorimer,~G.~H. Chaperonin-Mediated Protein Folding. \emph{Annu.
  Rev. Biophys. Biomol. Str.} \textbf{2001}, \emph{30}, 245--269\relax
\mciteBstWouldAddEndPuncttrue
\mciteSetBstMidEndSepPunct{\mcitedefaultmidpunct}
{\mcitedefaultendpunct}{\mcitedefaultseppunct}\relax
\EndOfBibitem
\bibitem[Matysiak \latin{et~al.}(2011)Matysiak, Debenedetti, and
  Rossky]{Rossky:2011:JPCB}
Matysiak,~S.; Debenedetti,~P.~G.; Rossky,~P.~J. Dissecting the Energetics of
  Hydrophobic Hydration of Polypeptides. \emph{J. Phys. Chem. B.}
  \textbf{2011}, \emph{115}, 14859--14865\relax
\mciteBstWouldAddEndPuncttrue
\mciteSetBstMidEndSepPunct{\mcitedefaultmidpunct}
{\mcitedefaultendpunct}{\mcitedefaultseppunct}\relax
\EndOfBibitem
\bibitem[Carmichael and Shell(2012)Carmichael, and Shell]{Shell:2012:JPCB}
Carmichael,~S.~P.; Shell,~M.~S. A New Multiscale Algorithm and Its Application
  to Coarse-Grained Peptide Models for Self-Assembly. \emph{J. Phys. Chem. B.}
  \textbf{2012}, \emph{116}, 8383--8393\relax
\mciteBstWouldAddEndPuncttrue
\mciteSetBstMidEndSepPunct{\mcitedefaultmidpunct}
{\mcitedefaultendpunct}{\mcitedefaultseppunct}\relax
\EndOfBibitem
\bibitem[{Bellissent-Funel} \latin{et~al.}(2016){Bellissent-Funel}, Hassanali,
  Havenith, Henchman, Pohl, Sterpone, {van der Spoel}, Xu, and
  Garcia]{bellissent2016water}
{Bellissent-Funel},~M.-C.; Hassanali,~A.; Havenith,~M.; Henchman,~R.; Pohl,~P.;
  Sterpone,~F.; {van der Spoel},~D.; Xu,~Y.; Garcia,~A.~E. Water Determines the
  Structure and Dynamics of Proteins. \emph{Chem. Rev.} \textbf{2016},
  \emph{116}, 7673--7697\relax
\mciteBstWouldAddEndPuncttrue
\mciteSetBstMidEndSepPunct{\mcitedefaultmidpunct}
{\mcitedefaultendpunct}{\mcitedefaultseppunct}\relax
\EndOfBibitem
\bibitem[DuBay \latin{et~al.}(2015)DuBay, Bowman, and
  Geissler]{Geissler:2015:ACR}
DuBay,~K.~H.; Bowman,~G.~R.; Geissler,~P.~L. Fluctuations within Folded
  Proteins: Implications for Thermodynamic and Allosteric Regulation.
  \emph{Acc. Chem. Res.} \textbf{2015}, \emph{48}, 1098--1105\relax
\mciteBstWouldAddEndPuncttrue
\mciteSetBstMidEndSepPunct{\mcitedefaultmidpunct}
{\mcitedefaultendpunct}{\mcitedefaultseppunct}\relax
\EndOfBibitem
\bibitem[Peter \latin{et~al.}(2016)Peter, Shea, and Pivkin]{Shea:2016:PCCP}
Peter,~E.~K.; Shea,~J.-E.; Pivkin,~I.~V. Coarse kMC-based replica exchange
  algorithms for the accelerated simulation of protein folding in explicit
  solvent. \emph{Phys. Chem. Chem. Phys.} \textbf{2016}, \emph{18},
  13052--13065\relax
\mciteBstWouldAddEndPuncttrue
\mciteSetBstMidEndSepPunct{\mcitedefaultmidpunct}
{\mcitedefaultendpunct}{\mcitedefaultseppunct}\relax
\EndOfBibitem
\bibitem[Lawrence \latin{et~al.}(2014)Lawrence, Kumar, Noid, and
  Showalter]{Noid:2014:JPCL}
Lawrence,~C.~W.; Kumar,~S.; Noid,~W.~G.; Showalter,~S.~A. Role of Ordered
  Proteins in the Folding-Upon-Binding of Intrinsically Disordered Proteins.
  \emph{J. Phys. Chem. Lett.} \textbf{2014}, \emph{5}, 833--838\relax
\mciteBstWouldAddEndPuncttrue
\mciteSetBstMidEndSepPunct{\mcitedefaultmidpunct}
{\mcitedefaultendpunct}{\mcitedefaultseppunct}\relax
\EndOfBibitem
\bibitem[Ferguson \latin{et~al.}(2010)Ferguson, Panagiotopoulos, Debenedetti,
  and Kevrekidis]{ferguson2010systematic}
Ferguson,~A.~L.; Panagiotopoulos,~A.~Z.; Debenedetti,~P.~G.; Kevrekidis,~I.~G.
  Systematic determination of order parameters for chain dynamics using
  diffusion maps. \emph{Proc. Natl. Acad. Sci. U.S.A.} \textbf{2010},
  \emph{107}, 13597--13602\relax
\mciteBstWouldAddEndPuncttrue
\mciteSetBstMidEndSepPunct{\mcitedefaultmidpunct}
{\mcitedefaultendpunct}{\mcitedefaultseppunct}\relax
\EndOfBibitem
\bibitem[ten Wolde and Chandler(2002)ten Wolde, and Chandler]{pc02}
ten Wolde,~P.~R.; Chandler,~D. Drying-induced hydrophobic polymer collapse.
  \emph{Proc. Natl. Acad. Sci. U.S.A.} \textbf{2002}, \emph{99},
  6539--6543\relax
\mciteBstWouldAddEndPuncttrue
\mciteSetBstMidEndSepPunct{\mcitedefaultmidpunct}
{\mcitedefaultendpunct}{\mcitedefaultseppunct}\relax
\EndOfBibitem
\bibitem[Miller \latin{et~al.}(2007)Miller, Vanden-Eijnden, and
  Chandler]{mvc07}
Miller,~T.~F.; Vanden-Eijnden,~E.; Chandler,~D. Solvent coarse-graining and the
  string method applied to the hydrophobic collapse of a hydrated chain.
  \emph{Proc. Natl. Acad. Sci. U.S.A.} \textbf{2007}, \emph{104},
  14559--14564\relax
\mciteBstWouldAddEndPuncttrue
\mciteSetBstMidEndSepPunct{\mcitedefaultmidpunct}
{\mcitedefaultendpunct}{\mcitedefaultseppunct}\relax
\EndOfBibitem
\bibitem[Berne \latin{et~al.}(2009)Berne, Weeks, and Zhou]{bwr09}
Berne,~B.~J.; Weeks,~J.~D.; Zhou,~R. Dewetting and Hydrophobic Interaction in
  Physical and Biological Systems. \emph{Annu. Rev. Phys. Chem.} \textbf{2009},
  \emph{60}, 85--103\relax
\mciteBstWouldAddEndPuncttrue
\mciteSetBstMidEndSepPunct{\mcitedefaultmidpunct}
{\mcitedefaultendpunct}{\mcitedefaultseppunct}\relax
\EndOfBibitem
\bibitem[Setny \latin{et~al.}(2013)Setny, Baron, Michael Kekenes-Huskey,
  McCammon, and Dzubiella]{Setny1197}
Setny,~P.; Baron,~R.; Michael Kekenes-Huskey,~P.; McCammon,~J.~A.;
  Dzubiella,~J. Solvent fluctuations in hydrophobic cavity{\textendash}ligand
  binding kinetics. \emph{Proc. Natl. Acad. Sci. U.S.A.} \textbf{2013},
  \emph{110}, 1197--1202\relax
\mciteBstWouldAddEndPuncttrue
\mciteSetBstMidEndSepPunct{\mcitedefaultmidpunct}
{\mcitedefaultendpunct}{\mcitedefaultseppunct}\relax
\EndOfBibitem
\bibitem[Remsing \latin{et~al.}(2015)Remsing, Xi, Vembanur, Sumit, Debenedetti,
  Garde, and Patel]{rxvsdg15}
Remsing,~R.~C.; Xi,~E.; Vembanur,~S.; Sumit,~S.; Debenedetti,~P.~G.; Garde,~S.;
  Patel,~A.~J. Pathways to dewetting in hydrophobic confinement. \emph{Proc.
  Natl. Acad. Sci. U.S.A.} \textbf{2015}, \emph{8181-8186}, 112\relax
\mciteBstWouldAddEndPuncttrue
\mciteSetBstMidEndSepPunct{\mcitedefaultmidpunct}
{\mcitedefaultendpunct}{\mcitedefaultseppunct}\relax
\EndOfBibitem
\bibitem[Tiwary \latin{et~al.}(2015)Tiwary, Mondal, Morrone, and
  Berne]{tiwary2015role}
Tiwary,~P.; Mondal,~J.; Morrone,~J.~A.; Berne,~B. Role of Water and Steric
  Constraints in the Kinetics of Cavity--Ligand Unbinding. \emph{Proc. Natl.
  Acad. Sci. U.S.A.} \textbf{2015}, \emph{112}, 12015--12019\relax
\mciteBstWouldAddEndPuncttrue
\mciteSetBstMidEndSepPunct{\mcitedefaultmidpunct}
{\mcitedefaultendpunct}{\mcitedefaultseppunct}\relax
\EndOfBibitem
\bibitem[Lum \latin{et~al.}(1999)Lum, Chandler, and Weeks]{lcw99}
Lum,~K.; Chandler,~D.; Weeks,~J.~D. Hydrophobicity at small and large length
  scales. \emph{J. Phys. Chem. B} \textbf{1999}, \emph{103}, 4570--4577\relax
\mciteBstWouldAddEndPuncttrue
\mciteSetBstMidEndSepPunct{\mcitedefaultmidpunct}
{\mcitedefaultendpunct}{\mcitedefaultseppunct}\relax
\EndOfBibitem
\bibitem[Athawale \latin{et~al.}(2007)Athawale, Goel, Ghosh, Truekett, and
  Garde]{garde07}
Athawale,~M.~V.; Goel,~G.; Ghosh,~T.; Truekett,~T.~M.; Garde,~S. Effects of
  lengthscales and attractions on the collapse of hydrophobic polymers in
  water. \emph{Proc. Natl. Acad. Sci. U.S.A.} \textbf{2007}, \emph{104},
  733--738\relax
\mciteBstWouldAddEndPuncttrue
\mciteSetBstMidEndSepPunct{\mcitedefaultmidpunct}
{\mcitedefaultendpunct}{\mcitedefaultseppunct}\relax
\EndOfBibitem
\bibitem[Goel \latin{et~al.}(2008)Goel, Athawale, Garde, and Truskett]{garde08}
Goel,~G.; Athawale,~M.~V.; Garde,~S.; Truskett,~T.~M. Attractions, Water
  Structure, and Thermodynamics of Hydrophobic Polymer Collapse. \emph{J. Phys.
  Chem. B} \textbf{2008}, \emph{112}, 13193--13196\relax
\mciteBstWouldAddEndPuncttrue
\mciteSetBstMidEndSepPunct{\mcitedefaultmidpunct}
{\mcitedefaultendpunct}{\mcitedefaultseppunct}\relax
\EndOfBibitem
\bibitem[Xi \latin{et~al.}(2018)Xi, Marks, Fialoke, and Patel]{xmfp18}
Xi,~E.; Marks,~S.~M.; Fialoke,~S.; Patel,~A.~J. Sparse sampling of water
  density fluctuations near liquid-vapor coexistence. \emph{Mol. Simul.}
  \textbf{2018}, \emph{44}, 1124--1135\relax
\mciteBstWouldAddEndPuncttrue
\mciteSetBstMidEndSepPunct{\mcitedefaultmidpunct}
{\mcitedefaultendpunct}{\mcitedefaultseppunct}\relax
\EndOfBibitem
\bibitem[Athawale \latin{et~al.}(2008)Athawale, Sarupria, and
  Garde]{garde08salt}
Athawale,~M.~V.; Sarupria,~S.; Garde,~S. Enthalpy-Entropy Contributions to Salt
  and Osmolyte Effects on Molecular-Scale Hydrophobic Hydration and
  Interactions. \emph{J. Phys. Chem. B} \textbf{2008}, \emph{112},
  5661--5670\relax
\mciteBstWouldAddEndPuncttrue
\mciteSetBstMidEndSepPunct{\mcitedefaultmidpunct}
{\mcitedefaultendpunct}{\mcitedefaultseppunct}\relax
\EndOfBibitem
\bibitem[Stirnemann \latin{et~al.}(2014)Stirnemann, Kang, Zhou, and
  Berne]{Stirnemann3413}
Stirnemann,~G.; Kang,~S.-g.; Zhou,~R.; Berne,~B.~J. How force unfolding differs
  from chemical denaturation. \emph{Proc. Natl. Acad. Sci. U.S.A.}
  \textbf{2014}, \emph{111}, 3413--3418\relax
\mciteBstWouldAddEndPuncttrue
\mciteSetBstMidEndSepPunct{\mcitedefaultmidpunct}
{\mcitedefaultendpunct}{\mcitedefaultseppunct}\relax
\EndOfBibitem
\bibitem[Mondal \latin{et~al.}(2015)Mondal, Halverson, Li, Stirnemann, Walker,
  and Berne]{mhlswb15}
Mondal,~J.; Halverson,~D.; Li,~I. T.~S.; Stirnemann,~G.; Walker,~G.~C.;
  Berne,~B. How osmolytes influence hydrophobic polymer conformations: A
  unified view from experiment and theory. \emph{Proc. Natl. Acad. Sci. U.S.A.}
  \textbf{2015}, \emph{112}, 9270--9275\relax
\mciteBstWouldAddEndPuncttrue
\mciteSetBstMidEndSepPunct{\mcitedefaultmidpunct}
{\mcitedefaultendpunct}{\mcitedefaultseppunct}\relax
\EndOfBibitem
\bibitem[Nayar and van~der Vegt(2018)Nayar, and van~der Vegt]{nv18}
Nayar,~D.; van~der Vegt,~N. F.~A. Cosolvent Effects on Polymer Hydration Drive
  Hydrophobic Collapse. \emph{J. Phys. Chem. B.} \textbf{2018}, \emph{122},
  3587--3595\relax
\mciteBstWouldAddEndPuncttrue
\mciteSetBstMidEndSepPunct{\mcitedefaultmidpunct}
{\mcitedefaultendpunct}{\mcitedefaultseppunct}\relax
\EndOfBibitem
\bibitem[Jamadagni \latin{et~al.}(2009)Jamadagni, Godawat, Dordick, and
  Garde]{Jamadagni-2009}
Jamadagni,~S.~N.; Godawat,~R.; Dordick,~J.~S.; Garde,~S. How Interfaces Affect
  Hydrophobically Driven Polymer Folding. \emph{J. Phys. Chem. B}
  \textbf{2009}, \emph{113}, 4093--4101, PMID: 19425248\relax
\mciteBstWouldAddEndPuncttrue
\mciteSetBstMidEndSepPunct{\mcitedefaultmidpunct}
{\mcitedefaultendpunct}{\mcitedefaultseppunct}\relax
\EndOfBibitem
\bibitem[Jamadagni \latin{et~al.}(2009)Jamadagni, Godawat, and
  Garde]{la9011839}
Jamadagni,~S.~N.; Godawat,~R.; Garde,~S. How Surface Wettability Affects the
  Binding, Folding, and Dynamics of Hydrophobic Polymers at Interfaces.
  \emph{Langmuir} \textbf{2009}, \emph{25}, 13092--13099\relax
\mciteBstWouldAddEndPuncttrue
\mciteSetBstMidEndSepPunct{\mcitedefaultmidpunct}
{\mcitedefaultendpunct}{\mcitedefaultseppunct}\relax
\EndOfBibitem
\bibitem[Vembanur \latin{et~al.}(2013)Vembanur, Patel, Sarupria, and
  Garde]{vpsg13}
Vembanur,~S.; Patel,~A.~J.; Sarupria,~S.; Garde,~S. On the Thermodynamics and
  Kinetics of Hydrophobic Interactions at Interfaces. \emph{J. Phys. Chem. B}
  \textbf{2013}, \emph{117}, 10261--10270\relax
\mciteBstWouldAddEndPuncttrue
\mciteSetBstMidEndSepPunct{\mcitedefaultmidpunct}
{\mcitedefaultendpunct}{\mcitedefaultseppunct}\relax
\EndOfBibitem
\bibitem[Zerze \latin{et~al.}(2015)Zerze, Mullen, Levine, Shea, and
  Mittal]{Shea:2015:Langmuir}
Zerze,~G.~H.; Mullen,~R.~G.; Levine,~Z.~A.; Shea,~J.-E.; Mittal,~J. To What
  Extent Does Surface Hydrophobicity Dictate Peptide Folding and Stability near
  Surfaces? \emph{Langmuir} \textbf{2015}, \emph{31}, 12223--12230\relax
\mciteBstWouldAddEndPuncttrue
\mciteSetBstMidEndSepPunct{\mcitedefaultmidpunct}
{\mcitedefaultendpunct}{\mcitedefaultseppunct}\relax
\EndOfBibitem
\bibitem[Jiang \latin{et~al.}(2019)Jiang, Remsing, Rego, and Patel]{jrrp19}
Jiang,~Z.; Remsing,~R.~C.; Rego,~N.~B.; Patel,~A.~J. Characterizing Solvent
  Density Fluctuations in Dynamical Observation Volumes. \emph{J. Phys. Chem.
  B} \textbf{2019}, \emph{123}, 1650--1661\relax
\mciteBstWouldAddEndPuncttrue
\mciteSetBstMidEndSepPunct{\mcitedefaultmidpunct}
{\mcitedefaultendpunct}{\mcitedefaultseppunct}\relax
\EndOfBibitem
\bibitem[Patel \latin{et~al.}(2012)Patel, Varilly, Jamadagni, Hagan, and
  Chandler]{pvjhc12}
Patel,~A.~J.; Varilly,~P.; Jamadagni,~S.~N.; Hagan,~M.~F.; Chandler,~D. Sitting
  at the Edge: How Biomolecules use Hydrophobicity to Tune Their Interactions
  and Function. \emph{J. Phys. Chem. B} \textbf{2012}, \emph{116},
  2498--2503\relax
\mciteBstWouldAddEndPuncttrue
\mciteSetBstMidEndSepPunct{\mcitedefaultmidpunct}
{\mcitedefaultendpunct}{\mcitedefaultseppunct}\relax
\EndOfBibitem
\bibitem[Patel and Garde(2014)Patel, and Garde]{pg14}
Patel,~A.~J.; Garde,~S. Efficient Method To Characterize the Context-Dependent
  Hydrophobicity of Proteins. \emph{J. Phys. Chem. B} \textbf{2014},
  \emph{118}, 1564--1573\relax
\mciteBstWouldAddEndPuncttrue
\mciteSetBstMidEndSepPunct{\mcitedefaultmidpunct}
{\mcitedefaultendpunct}{\mcitedefaultseppunct}\relax
\EndOfBibitem
\bibitem[Patel \latin{et~al.}(2010)Patel, Varilly, and Chandler]{pvc10}
Patel,~A.~J.; Varilly,~P.; Chandler,~D. Fluctuations of Water near Extended
  Hydrophobic and Hydrophilic Surfaces. \emph{J. Phys. Chem. B} \textbf{2010},
  \emph{114}, 1632--1637\relax
\mciteBstWouldAddEndPuncttrue
\mciteSetBstMidEndSepPunct{\mcitedefaultmidpunct}
{\mcitedefaultendpunct}{\mcitedefaultseppunct}\relax
\EndOfBibitem
\bibitem[Patel \latin{et~al.}(2011)Patel, Varilly, and Chandler]{pvc11}
Patel,~A.~J.; Varilly,~P.; Chandler,~D. Quantifying Density Fluctuations in
  Volumes of All Shapes and Sizes Using Indirect Umbrella Sampling. \emph{J.
  Stat. Phys.} \textbf{2011}, \emph{145}, 265--275\relax
\mciteBstWouldAddEndPuncttrue
\mciteSetBstMidEndSepPunct{\mcitedefaultmidpunct}
{\mcitedefaultendpunct}{\mcitedefaultseppunct}\relax
\EndOfBibitem
\bibitem[Souaille and Roux(2001)Souaille, and Roux]{roux}
Souaille,~M.; Roux,~B. Extension to the weighted histogram analysis method:
  combining umbrella sampling with free energy calculations. \emph{Comput.
  Phys. Commun.} \textbf{2001}, \emph{135}, 40--57\relax
\mciteBstWouldAddEndPuncttrue
\mciteSetBstMidEndSepPunct{\mcitedefaultmidpunct}
{\mcitedefaultendpunct}{\mcitedefaultseppunct}\relax
\EndOfBibitem
\bibitem[Shirts and Chodera(2008)Shirts, and Chodera]{mbar}
Shirts,~M.~R.; Chodera,~J.~D. Statistically optimal analysis of samples from
  multiple equilibrium states. \emph{J. Chem. Phys.} \textbf{2008}, \emph{129},
  124105\relax
\mciteBstWouldAddEndPuncttrue
\mciteSetBstMidEndSepPunct{\mcitedefaultmidpunct}
{\mcitedefaultendpunct}{\mcitedefaultseppunct}\relax
\EndOfBibitem
\bibitem[Zhu and Hummer(2012)Zhu, and Hummer]{zh12}
Zhu,~F.; Hummer,~G. Convergence and error estimation in free energy
  calculations using the weighted histogram analysis method. \emph{J. Comput.
  Chem.} \textbf{2012}, \emph{33}, 453--465\relax
\mciteBstWouldAddEndPuncttrue
\mciteSetBstMidEndSepPunct{\mcitedefaultmidpunct}
{\mcitedefaultendpunct}{\mcitedefaultseppunct}\relax
\EndOfBibitem
\bibitem[Tan \latin{et~al.}(2012)Tan, Gallichio, Lapelosa, and Levy]{UWHAM}
Tan,~Z.; Gallichio,~E.; Lapelosa,~M.; Levy,~R.~M. Theory of binless multi-state
  free energy estimation with applications to protein-ligand binding. \emph{J.
  Chem. Phys.} \textbf{2012}, \emph{136}, 144102\relax
\mciteBstWouldAddEndPuncttrue
\mciteSetBstMidEndSepPunct{\mcitedefaultmidpunct}
{\mcitedefaultendpunct}{\mcitedefaultseppunct}\relax
\EndOfBibitem
\bibitem[Xi \latin{et~al.}(2016)Xi, Remsing, and Patel]{xrp16}
Xi,~E.; Remsing,~R.~C.; Patel,~A.~J. Sparse Sampling of Water Density
  Fluctuations in Interfacial Environments. \emph{J. Chem. Theory Comput.}
  \textbf{2016}, \emph{12}, 706--713\relax
\mciteBstWouldAddEndPuncttrue
\mciteSetBstMidEndSepPunct{\mcitedefaultmidpunct}
{\mcitedefaultendpunct}{\mcitedefaultseppunct}\relax
\EndOfBibitem
\bibitem[Hess \latin{et~al.}(2008)Hess, Kutzner, van~der Spoel, and
  Lindahl]{hkvl08}
Hess,~B.; Kutzner,~C.; van~der Spoel,~D.; Lindahl,~E. GROMACS 4: Algorithms for
  Highly Efficient, Load-Balanced, and Scalable Molecular Simulation. \emph{J.
  Chem. Theory Comput.} \textbf{2008}, \emph{4}, 435--447\relax
\mciteBstWouldAddEndPuncttrue
\mciteSetBstMidEndSepPunct{\mcitedefaultmidpunct}
{\mcitedefaultendpunct}{\mcitedefaultseppunct}\relax
\EndOfBibitem
\bibitem[Bussi \latin{et~al.}(2007)Bussi, Donadio, and
  Parrinello]{Bussi:JCP:2007}
Bussi,~G.; Donadio,~D.; Parrinello,~M. Canonical Sampling through Velocity
  Rescaling. \emph{J. Chem. Phys.} \textbf{2007}, \emph{126}, 014101\relax
\mciteBstWouldAddEndPuncttrue
\mciteSetBstMidEndSepPunct{\mcitedefaultmidpunct}
{\mcitedefaultendpunct}{\mcitedefaultseppunct}\relax
\EndOfBibitem
\bibitem[Parrinello and Rahman(1981)Parrinello, and Rahman]{Parrinello-Rahman}
Parrinello,~M.; Rahman,~A. Polymorphic Transitions in Single Crystals: A New
  Molecular Dynamics Method. \emph{J. Appl. Phys.} \textbf{1981}, \emph{52},
  7182--7190\relax
\mciteBstWouldAddEndPuncttrue
\mciteSetBstMidEndSepPunct{\mcitedefaultmidpunct}
{\mcitedefaultendpunct}{\mcitedefaultseppunct}\relax
\EndOfBibitem
\bibitem[Martin and Siepmann(1998)Martin, and Siepmann]{ms98}
Martin,~M.~G.; Siepmann,~J.~I. Transferable Potentials for Phase Equilibria. 1.
  United-Atom Description of n-Alkanes. \emph{J. Phys. Chem. B.} \textbf{1998},
  \emph{102}, 2569--2577\relax
\mciteBstWouldAddEndPuncttrue
\mciteSetBstMidEndSepPunct{\mcitedefaultmidpunct}
{\mcitedefaultendpunct}{\mcitedefaultseppunct}\relax
\EndOfBibitem
\bibitem[Berendsen \latin{et~al.}(1987)Berendsen, Grigera, and
  Straatsma]{bgs87}
Berendsen,~H. J.~C.; Grigera,~J.~R.; Straatsma,~T.~P. The Missing Term in
  Effective Pair Potentials. \emph{J. Phys. Chem.} \textbf{1987}, \emph{91},
  6269--6271\relax
\mciteBstWouldAddEndPuncttrue
\mciteSetBstMidEndSepPunct{\mcitedefaultmidpunct}
{\mcitedefaultendpunct}{\mcitedefaultseppunct}\relax
\EndOfBibitem
\bibitem[Miyamoto and Kollman(1992)Miyamoto, and Kollman]{SETTLE}
Miyamoto,~S.; Kollman,~P.~A. Settle: An analytical version of the SHAKE and
  RATTLE algorithm for rigid water models. \emph{J. Comput. Chem.}
  \textbf{1992}, \emph{13}, 952--962\relax
\mciteBstWouldAddEndPuncttrue
\mciteSetBstMidEndSepPunct{\mcitedefaultmidpunct}
{\mcitedefaultendpunct}{\mcitedefaultseppunct}\relax
\EndOfBibitem
\bibitem[Essmann \latin{et~al.}(1995)Essmann, Perera, Berkowitz, Darden, Lee,
  and Pedersen]{PME}
Essmann,~U.; Perera,~L.; Berkowitz,~M.~L.; Darden,~T.; Lee,~H.; Pedersen,~L.~G.
  A Smooth Particle Mesh Ewald Method. \emph{J. Chem. Phys.} \textbf{1995},
  \emph{103}, 8577--8593\relax
\mciteBstWouldAddEndPuncttrue
\mciteSetBstMidEndSepPunct{\mcitedefaultmidpunct}
{\mcitedefaultendpunct}{\mcitedefaultseppunct}\relax
\EndOfBibitem
\bibitem[Rego \latin{et~al.}(2019)Rego, Xi, and Patel]{rxp19}
Rego,~N.~B.; Xi,~E.; Patel,~A.~J. Protein Hydration Waters are Susceptible to
  Unfavourable Perturbations. \emph{J. Am. Chem. Soc.} \textbf{2019},
  \emph{141}, 2080--2086\relax
\mciteBstWouldAddEndPuncttrue
\mciteSetBstMidEndSepPunct{\mcitedefaultmidpunct}
{\mcitedefaultendpunct}{\mcitedefaultseppunct}\relax
\EndOfBibitem
\bibitem[Remsing and Patel(2015)Remsing, and Patel]{rp15}
Remsing,~R.~C.; Patel,~A.~J. Water density fluctuations relevant to hydrophobic
  hydration are unaltered by attractions. \emph{J. Chem. Phys.} \textbf{2015},
  \emph{142}, 024502\relax
\mciteBstWouldAddEndPuncttrue
\mciteSetBstMidEndSepPunct{\mcitedefaultmidpunct}
{\mcitedefaultendpunct}{\mcitedefaultseppunct}\relax
\EndOfBibitem
\bibitem[Wu and Garde(2015)Wu, and Garde]{wg15}
Wu,~E.; Garde,~S. Lengthscale-Dependent Solvation and Density Fluctuations in
  n-Octane. \emph{J. Phys. Chem. B} \textbf{2015}, \emph{119}, 9287--9294\relax
\mciteBstWouldAddEndPuncttrue
\mciteSetBstMidEndSepPunct{\mcitedefaultmidpunct}
{\mcitedefaultendpunct}{\mcitedefaultseppunct}\relax
\EndOfBibitem
\bibitem[Li and Walker(2011)Li, and Walker]{Li-Walker}
Li,~I. T.~S.; Walker,~G.~C. Signature of hydrophobic hydration in a single
  polymer. \emph{Proc. Natl. Acad. Sci. U.S.A.} \textbf{2011}, \emph{108},
  16527--16532\relax
\mciteBstWouldAddEndPuncttrue
\mciteSetBstMidEndSepPunct{\mcitedefaultmidpunct}
{\mcitedefaultendpunct}{\mcitedefaultseppunct}\relax
\EndOfBibitem
\bibitem[Garde and Patel(2011)Garde, and Patel]{gp11}
Garde,~S.; Patel,~A.~J. Unraveling the hydrophobic effect, one molecule at a
  time. \emph{Proc. Natl. Acad. Sci. U.S.A.} \textbf{2011}, \emph{108},
  16491--16492\relax
\mciteBstWouldAddEndPuncttrue
\mciteSetBstMidEndSepPunct{\mcitedefaultmidpunct}
{\mcitedefaultendpunct}{\mcitedefaultseppunct}\relax
\EndOfBibitem
\bibitem[Hummer \latin{et~al.}(1996)Hummer, Garde, Garcia, Pohorille, and
  Pratt]{hggpp96}
Hummer,~G.; Garde,~S.; Garcia,~A.~E.; Pohorille,~A.; Pratt,~L.~R. An
  Information Theory Model of Hydrophobic Interactions. \emph{Proc. Natl. Acad.
  Sci. U.S.A.} \textbf{1996}, \emph{93}, 8951--8955\relax
\mciteBstWouldAddEndPuncttrue
\mciteSetBstMidEndSepPunct{\mcitedefaultmidpunct}
{\mcitedefaultendpunct}{\mcitedefaultseppunct}\relax
\EndOfBibitem
\bibitem[Rajamani \latin{et~al.}(2005)Rajamani, Truskett, and Garde]{rts05}
Rajamani,~S.; Truskett,~T.~M.; Garde,~S. Hydrophobic hydration from small to
  large lengthscales: Understanding and manipulating the crossover. \emph{Proc.
  Natl. Acad. Sci. U.S.A.} \textbf{2005}, \emph{102}, 9475--9480\relax
\mciteBstWouldAddEndPuncttrue
\mciteSetBstMidEndSepPunct{\mcitedefaultmidpunct}
{\mcitedefaultendpunct}{\mcitedefaultseppunct}\relax
\EndOfBibitem
\bibitem[Varilly \latin{et~al.}(2011)Varilly, Patel, and Chandler]{vpc11}
Varilly,~P.; Patel,~A.~J.; Chandler,~D. An improved coarse-grained model of
  solvation and the hydrophobic effect. \emph{J. Chem. Phys.} \textbf{2011},
  \emph{134}, 074109\relax
\mciteBstWouldAddEndPuncttrue
\mciteSetBstMidEndSepPunct{\mcitedefaultmidpunct}
{\mcitedefaultendpunct}{\mcitedefaultseppunct}\relax
\EndOfBibitem
\bibitem[Vaikuntanathan \latin{et~al.}(2016)Vaikuntanathan, Rotskoff, Hudson,
  and Geissler]{Geissler:2016:PNAS}
Vaikuntanathan,~S.; Rotskoff,~G.; Hudson,~A.; Geissler,~P.~L. Necessity of
  Capillary Modes in a Minimal Model of Nanoscale Hydrophobic Solvation.
  \emph{Proc. Natl. Acad. Sci. U.S.A.} \textbf{2016}, \emph{113},
  E2224--E2230\relax
\mciteBstWouldAddEndPuncttrue
\mciteSetBstMidEndSepPunct{\mcitedefaultmidpunct}
{\mcitedefaultendpunct}{\mcitedefaultseppunct}\relax
\EndOfBibitem
\bibitem[Xi and Patel(2016)Xi, and Patel]{xp16}
Xi,~E.; Patel,~A.~J. The hydrophobic effect, and fluctuations: the long and the
  short of it. \emph{Proc. Natl. Acad. Sci. U.S.A.} \textbf{2016}, \emph{113},
  4549--4551\relax
\mciteBstWouldAddEndPuncttrue
\mciteSetBstMidEndSepPunct{\mcitedefaultmidpunct}
{\mcitedefaultendpunct}{\mcitedefaultseppunct}\relax
\EndOfBibitem
\bibitem[Huang and Chandler(2000)Huang, and Chandler]{Huang:2000ve}
Huang,~D.~M.; Chandler,~D. Cavity formation and the drying transition in the
  Lennard-Jones fluid. \emph{Phys. Rev. E} \textbf{2000}, \emph{61},
  1501--1506\relax
\mciteBstWouldAddEndPuncttrue
\mciteSetBstMidEndSepPunct{\mcitedefaultmidpunct}
{\mcitedefaultendpunct}{\mcitedefaultseppunct}\relax
\EndOfBibitem
\bibitem[Cerdeirina \latin{et~al.}(2011)Cerdeirina, Debenedetti, Rossky, and
  Giovambattista]{pgd:JPCL:2011}
Cerdeirina,~C.~A.; Debenedetti,~P.~G.; Rossky,~P.~J.; Giovambattista,~N.
  Evaporation Length Scales of Confined Water and Some Common Organic Liquids.
  \emph{J. Phys. Chem. Lett.} \textbf{2011}, \emph{2}, 1000--1003\relax
\mciteBstWouldAddEndPuncttrue
\mciteSetBstMidEndSepPunct{\mcitedefaultmidpunct}
{\mcitedefaultendpunct}{\mcitedefaultseppunct}\relax
\EndOfBibitem
\bibitem[Garde \latin{et~al.}(1996)Garde, Hummer, Garica, Paulaitis, and
  Pratt]{ghgpp96}
Garde,~S.; Hummer,~G.; Garica,~A.~E.; Paulaitis,~M.~E.; Pratt,~L.~R. Origin of
  Entropy Convergence in Hydrophobic Hydration and Protein Folding. \emph{Phys.
  Rev. Lett.} \textbf{1996}, \emph{77}, 4966\relax
\mciteBstWouldAddEndPuncttrue
\mciteSetBstMidEndSepPunct{\mcitedefaultmidpunct}
{\mcitedefaultendpunct}{\mcitedefaultseppunct}\relax
\EndOfBibitem
\bibitem[Godawat \latin{et~al.}(2009)Godawat, Jamadagni, and Garde]{gjg09}
Godawat,~R.; Jamadagni,~S.~N.; Garde,~S. Characterizing hydrophobicity of
  interfaces by using cavity formation, solute binding, and water correlations.
  \emph{Proc. Natl. Acad. Sci. U.S.A.} \textbf{2009}, \emph{106},
  15119--15124\relax
\mciteBstWouldAddEndPuncttrue
\mciteSetBstMidEndSepPunct{\mcitedefaultmidpunct}
{\mcitedefaultendpunct}{\mcitedefaultseppunct}\relax
\EndOfBibitem
\bibitem[Acharya \latin{et~al.}(2010)Acharya, Vembanur, Jamadagni, and
  Garde]{Acharya:Faraday:2010}
Acharya,~H.; Vembanur,~S.; Jamadagni,~S.~N.; Garde,~S. Mapping Hydrophobicity
  at the Nanoscale: {{Applications}} to Heterogeneous Surfaces and Proteins.
  \emph{Faraday Discuss.} \textbf{2010}, \emph{146}, 353--365\relax
\mciteBstWouldAddEndPuncttrue
\mciteSetBstMidEndSepPunct{\mcitedefaultmidpunct}
{\mcitedefaultendpunct}{\mcitedefaultseppunct}\relax
\EndOfBibitem
\bibitem[Rotenberg \latin{et~al.}(2011)Rotenberg, Patel, and
  Chandler]{Rotenberg:JACS:2011}
Rotenberg,~B.; Patel,~A.~J.; Chandler,~D. Molecular Explanation for Why Talc
  Surfaces can be Both Hydrophilic and Hydrophobic. \emph{J. Am. Chem. Soc.}
  \textbf{2011}, \emph{133}, 20521 -- 20527\relax
\mciteBstWouldAddEndPuncttrue
\mciteSetBstMidEndSepPunct{\mcitedefaultmidpunct}
{\mcitedefaultendpunct}{\mcitedefaultseppunct}\relax
\EndOfBibitem
\bibitem[Jamadagni \latin{et~al.}(2011)Jamadagni, Godawat, and
  Garde]{Jamadagni:ARCB:2011}
Jamadagni,~S.~N.; Godawat,~R.; Garde,~S. Hydrophobicity of {{Proteins}} and
  {{Interfaces}}: {{Insights}} from {{Density Fluctuations}}. \emph{Annu. Rev.
  Chem. Biomol. Eng.} \textbf{2011}, \emph{2}, 147--171\relax
\mciteBstWouldAddEndPuncttrue
\mciteSetBstMidEndSepPunct{\mcitedefaultmidpunct}
{\mcitedefaultendpunct}{\mcitedefaultseppunct}\relax
\EndOfBibitem
\bibitem[Maibaum \latin{et~al.}(2004)Maibaum, Dinner, and Chandler]{micelle}
Maibaum,~L.; Dinner,~A.~R.; Chandler,~D. Micelle Formation and the Hydrophobic
  Effect. \emph{J. Phys. Chem. B} \textbf{2004}, \emph{108}, 6778--6781\relax
\mciteBstWouldAddEndPuncttrue
\mciteSetBstMidEndSepPunct{\mcitedefaultmidpunct}
{\mcitedefaultendpunct}{\mcitedefaultseppunct}\relax
\EndOfBibitem
\bibitem[Patel \latin{et~al.}(2011)Patel, Varilly, Jamadagni, Acharya, Garde,
  and Chandler]{pvjagc11}
Patel,~A.~J.; Varilly,~P.; Jamadagni,~S.~N.; Acharya,~H.; Garde,~S.;
  Chandler,~D. Extended surfaces modulate hydrophobic interactions of
  neighboring solutes. \emph{Proc. Natl. Acad. Sci. U.S.A.} \textbf{2011},
  \emph{108}, 17678--17683\relax
\mciteBstWouldAddEndPuncttrue
\mciteSetBstMidEndSepPunct{\mcitedefaultmidpunct}
{\mcitedefaultendpunct}{\mcitedefaultseppunct}\relax
\EndOfBibitem
\bibitem[Prakash \latin{et~al.}(2016)Prakash, Xi, and Patel]{pxp16}
Prakash,~S.; Xi,~E.; Patel,~A.~J. Spontaneous recovery of superhydrophobicity
  on nanotextured surfaces. \emph{Proc. Natl. Acad. Sci. U.S.A.} \textbf{2016},
  \emph{113}, 5508--5513\relax
\mciteBstWouldAddEndPuncttrue
\mciteSetBstMidEndSepPunct{\mcitedefaultmidpunct}
{\mcitedefaultendpunct}{\mcitedefaultseppunct}\relax
\EndOfBibitem
\bibitem[Xi \latin{et~al.}(2017)Xi, Venkateshwaran, Li, Rego, Patel, and
  Garde]{xvlrpg14}
Xi,~E.; Venkateshwaran,~V.; Li,~L.; Rego,~N.; Patel,~A.~J.; Garde,~S.
  Hydrophobicity of proteins and nanostructured solutes is governed by
  topographical and chemical context. \emph{Proc. Natl. Acad. Sci. U.S.A.}
  \textbf{2017}, \emph{114}, 13345--13350\relax
\mciteBstWouldAddEndPuncttrue
\mciteSetBstMidEndSepPunct{\mcitedefaultmidpunct}
{\mcitedefaultendpunct}{\mcitedefaultseppunct}\relax
\EndOfBibitem
\bibitem[Rego \latin{et~al.}(2021)Rego, Xi, and Patel]{Rego:2021gx}
Rego,~N.~B.; Xi,~E.; Patel,~A.~J. Identifying hydrophobic protein patches to
  inform protein interaction interfaces. \emph{Proc. Natl. Acad. Sci. U.S.A.}
  \textbf{2021}, \emph{118}, e2018234118\relax
\mciteBstWouldAddEndPuncttrue
\mciteSetBstMidEndSepPunct{\mcitedefaultmidpunct}
{\mcitedefaultendpunct}{\mcitedefaultseppunct}\relax
\EndOfBibitem
\bibitem[Rodriguez-Ropero \latin{et~al.}(2015)Rodriguez-Ropero, Hajari, and
  van~der Vegt]{rhv15}
Rodriguez-Ropero,~F.; Hajari,~T.; van~der Vegt,~N. F.~A. Mechanism of Polymer
  Collapse in Miscible Good Solvents. \emph{J. Phys. Chem. B} \textbf{2015},
  \emph{119}, 15780--15788\relax
\mciteBstWouldAddEndPuncttrue
\mciteSetBstMidEndSepPunct{\mcitedefaultmidpunct}
{\mcitedefaultendpunct}{\mcitedefaultseppunct}\relax
\EndOfBibitem
\bibitem[Budkov \latin{et~al.}(2015)Budkov, Kolesnikov, Georgi, and
  Kiselev]{yanm15}
Budkov,~Y.~A.; Kolesnikov,~A.~L.; Georgi,~N.; Kiselev,~M.~G. A flexible polymer
  chain in a critical solvent: Coil or globule? \emph{Europhys. Lett.}
  \textbf{2015}, \emph{109}, 36005\relax
\mciteBstWouldAddEndPuncttrue
\mciteSetBstMidEndSepPunct{\mcitedefaultmidpunct}
{\mcitedefaultendpunct}{\mcitedefaultseppunct}\relax
\EndOfBibitem
\bibitem[Dignon \latin{et~al.}(2018)Dignon, Zheng, Best, Kim, and
  Mittal]{dzbkm18}
Dignon,~G.~L.; Zheng,~W.; Best,~R.~B.; Kim,~Y.~C.; Mittal,~J. Relation between
  single-molecule properties and phase behavior of intrinsically disordered
  proteins. \emph{Proc. Natl. Acad. Sci. U.S.A.} \textbf{2018}, \emph{115},
  9929--9934\relax
\mciteBstWouldAddEndPuncttrue
\mciteSetBstMidEndSepPunct{\mcitedefaultmidpunct}
{\mcitedefaultendpunct}{\mcitedefaultseppunct}\relax
\EndOfBibitem
\bibitem[Bharadwaj and van~der Vegt(2019)Bharadwaj, and van~der Vegt]{bv19}
Bharadwaj,~S.; van~der Vegt,~N. F.~A. Does Preferential Adsorption Drive
  Cononsolvency? \emph{Macromolecules} \textbf{2019}, \emph{52},
  4131--4138\relax
\mciteBstWouldAddEndPuncttrue
\mciteSetBstMidEndSepPunct{\mcitedefaultmidpunct}
{\mcitedefaultendpunct}{\mcitedefaultseppunct}\relax
\EndOfBibitem
\bibitem[Dubou{\'e}-Dijon \latin{et~al.}(2016)Dubou{\'e}-Dijon, Fogarty, Hynes,
  and Laage]{elise-dna}
Dubou{\'e}-Dijon,~E.; Fogarty,~A.~C.; Hynes,~J.~T.; Laage,~D. Dynamical
  Disorder in the DNA Hydration Shell. \emph{J. Am. Chem. Soc.} \textbf{2016},
  \emph{138}, 7610--7620\relax
\mciteBstWouldAddEndPuncttrue
\mciteSetBstMidEndSepPunct{\mcitedefaultmidpunct}
{\mcitedefaultendpunct}{\mcitedefaultseppunct}\relax
\EndOfBibitem
\bibitem[Ganguly \latin{et~al.}(2016)Ganguly, van~der Vegt, and
  Shea]{Shea:2016:JPCL}
Ganguly,~P.; van~der Vegt,~N. F.~A.; Shea,~J.-E. Hydrophobic Association in
  Mixed Urea--TMAO Solutions. \emph{J. Phys. Chem. Lett.} \textbf{2016},
  \emph{7}, 3052--3059\relax
\mciteBstWouldAddEndPuncttrue
\mciteSetBstMidEndSepPunct{\mcitedefaultmidpunct}
{\mcitedefaultendpunct}{\mcitedefaultseppunct}\relax
\EndOfBibitem
\bibitem[Nayar and van~der Vegt(2017)Nayar, and van~der Vegt]{nv17}
Nayar,~D.; van~der Vegt,~N. F.~A. The Hydrophobic Effect and the Role of
  Cosolvents. \emph{J. Phys. Chem. B} \textbf{2017}, \emph{121},
  9986--9998\relax
\mciteBstWouldAddEndPuncttrue
\mciteSetBstMidEndSepPunct{\mcitedefaultmidpunct}
{\mcitedefaultendpunct}{\mcitedefaultseppunct}\relax
\EndOfBibitem
\bibitem[Remsing \latin{et~al.}(2018)Remsing, Xi, and Patel]{Remsing:JPCB:2018}
Remsing,~R.~C.; Xi,~E.; Patel,~A.~J. Protein {{Hydration Thermodynamics}}:
  {{The Influence}} of {{Flexibility}} and {{Salt}} on {{Hydrophobin II
  Hydration}}. \emph{J. Phys. Chem. B} \textbf{2018}, \emph{122},
  3635--3646\relax
\mciteBstWouldAddEndPuncttrue
\mciteSetBstMidEndSepPunct{\mcitedefaultmidpunct}
{\mcitedefaultendpunct}{\mcitedefaultseppunct}\relax
\EndOfBibitem
\bibitem[Zhang and Cremer(2010)Zhang, and Cremer]{cremer-arpc}
Zhang,~Y.; Cremer,~P.~S. Chemistry of Hofmeister Anions and Osmolytes.
  \emph{Annu. Rev. Phys. Chem.} \textbf{2010}, \emph{61}, 63--83\relax
\mciteBstWouldAddEndPuncttrue
\mciteSetBstMidEndSepPunct{\mcitedefaultmidpunct}
{\mcitedefaultendpunct}{\mcitedefaultseppunct}\relax
\EndOfBibitem
\bibitem[Okur \latin{et~al.}(2017)Okur, Hlad{\'\i}lkov{\'a}, Rembert, Cho,
  Heyda, Dzubiella, Cremer, and Jungwirth]{okur2017beyond}
Okur,~H.~I.; Hlad{\'\i}lkov{\'a},~J.; Rembert,~K.~B.; Cho,~Y.; Heyda,~J.;
  Dzubiella,~J.; Cremer,~P.~S.; Jungwirth,~P. Beyond the {{Hofmeister}} series:
  ion-specific effects on proteins and their biological functions. \emph{J.
  Phys. Chem. B} \textbf{2017}, \emph{121}, 1997--2014\relax
\mciteBstWouldAddEndPuncttrue
\mciteSetBstMidEndSepPunct{\mcitedefaultmidpunct}
{\mcitedefaultendpunct}{\mcitedefaultseppunct}\relax
\EndOfBibitem
\bibitem[Mukherji \latin{et~al.}(2019)Mukherji, Watson, Morsbach, Schmutz,
  Wagner, Marques, and Kremer]{mwmswm19}
Mukherji,~D.; Watson,~M.~D.; Morsbach,~S.; Schmutz,~M.; Wagner,~M.;
  Marques,~C.~M.; Kremer,~K. Soft and Smart: Co-nonsolvency-Based Design of
  Multiresponsive Copolymers. \emph{Macromolecules} \textbf{2019}, \emph{52},
  3471--3478\relax
\mciteBstWouldAddEndPuncttrue
\mciteSetBstMidEndSepPunct{\mcitedefaultmidpunct}
{\mcitedefaultendpunct}{\mcitedefaultseppunct}\relax
\EndOfBibitem
\end{mcitethebibliography}

\providecommand{\latin}[1]{#1}
\makeatletter
\providecommand{\doi}
  {\begingroup\let\do\@makeother\dospecials
  \catcode`\{=1 \catcode`\}=2 \doi@aux}
\providecommand{\doi@aux}[1]{\endgroup\texttt{#1}}
\makeatother
\providecommand*\mcitethebibliography{\thebibliography}
\csname @ifundefined\endcsname{endmcitethebibliography}
  {\let\endmcitethebibliography\endthebibliography}{}

\includepdf[pages=1-last]{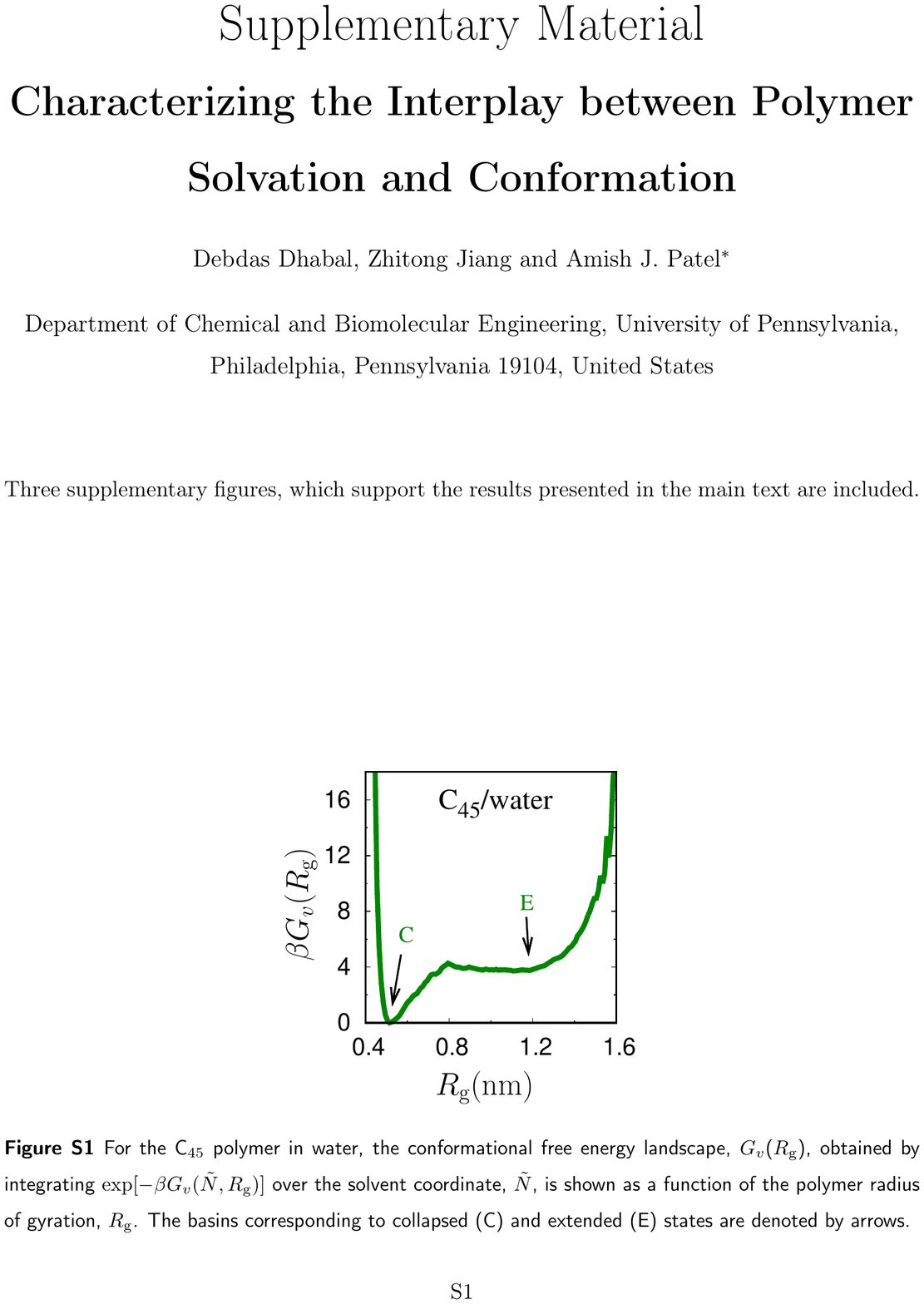}
\end{document}